\documentclass{article}
\usepackage[T1]{fontenc}
\usepackage[latin9]{inputenc}
\usepackage{geometry}
\usepackage{amsmath}
\usepackage{amssymb}
\usepackage{graphicx}
\usepackage{setspace}
\usepackage[authoryear]{natbib}

\makeatletter

\providecommand{\tabularnewline}{\\}

\usepackage{adjustbox}
\usepackage{multirow}
\usepackage{algorithm}
\usepackage{algpseudocode}%\usepackage{rotating}
\usepackage{pdflscape}
\usepackage{bibunits}
\usepackage{authblk}

\defaultbibliographystyle{chicago}
\defaultbibliography{HMC_BibTex}

\let\oldmaketitle\maketitle% Store old \maketitle
\renewcommand{\maketitle}{% Update \maketitle
\begin{singlespace}
  \oldmaketitle% Use singlespace
\end{singlespace}}

\makeatother

\title{Importance Sampling-based Transport Map Hamiltonian Monte Carlo for
Bayesian Hierarchical Models}
\author[1]{Kjartan Kloster Osmundsen\thanks{Corresponding author. Email: kjartan.osmundsen@gmail.com}}
\author[1]{Tore Selland Kleppe}
\author[2]{Roman Liesenfeld}
\affil[1]{Department of Mathematics and Physics, University of Stavanger, Norway}
\affil[2]{Institute of Econometrics and Statistics, University of Cologne,Germany}

\begin{document}
\begin{bibunit}

\maketitle

\begin{abstract}
We propose an importance sampling (IS)-based transport map Hamiltonian Monte Carlo procedure for performing full Bayesian analysis in general nonlinear high-dimensional hierarchical models. Using IS techniques to construct a transport map, the proposed method transforms the typically highly challenging target distribution of a hierarchical model into a target which is easily sampled using standard Hamiltonian Monte Carlo. Conventional applications of high-dimensional IS, where infinite variance of IS weights can be a serious problem, require computationally costly high-fidelity IS distributions. An appealing property of our method is that the IS distributions employed can be of rather low fidelity, making it computationally cheap. We illustrate our algorithm in applications to challenging dynamic state-space models, where it exhibits very high simulation efficiency compared to relevant benchmarks, even for variants of the proposed method implemented using a few dozen lines of code in the Stan statistical software.
\end{abstract}

\noindent
\begin{keywords} Hamiltonian Monte Carlo; Importance Sampling; Transport Map; Bayesian hierarchical models;  
State-space models; Stan \end{keywords}

\section{Introduction\label{intro}}

Computational methods for Bayesian nonlinear/non-Gaussian hierarchical
models is an active field of research, and advances in such computational
methods allow researchers to build and fit progressively more complex
models. Existing Markov chain Monte Carlo (MCMC) methods for such
models fall broadly into four categories. Firstly, Gibbs sampling
is widely used, in part due to its simple implementation \citep[see e.g.][]{Robert2004}.
However, a naive implementation updating latent variables in one block
and model parameters in another block can suffer from a very slow
exploration \citep[see e.g.][]{jacquier_etal94} of the target distribution if this joint distribution implies
a strong, typically nonlinear dependence structure of the variables
in the two blocks. Secondly, methods that update latent variables
and parameters jointly avoid the nonlinear dependence problem of
Gibbs sampling. One such approach for joint updates is to use Riemann
manifold Hamiltonian Monte Carlo (RMHMC) methods \citep[see e.g.][]{girolami_calderhead_11,NIPS2014_5591,Kleppe2018}.
However, they critically require update proposals which are properly
aligned with the (typically rather variable) local geometry of the
target, the generation of which can be computationally demanding for
complex high-dimensional joint posteriors of the parameters and latent
variables. 

The third category is pseudo-marginal methods \citep[see e.g.][and references therein]{Andrieu2010,Pitt2012134},
which bypasses the problematic parameters and latent variables dependency
by targeting directly the marginal posterior of the parameters. Pseudo-marginal methods
require, however, a low variance, unbiased Monte Carlo (MC) estimate
of said posterior, which can often be extremely computationally demanding for high-dimensional models \citep[see e.g.][]{flury_shephart_2011}. Moreover, for models with
many parameters, it can be difficult to select an efficient proposal
distribution for updating the parameters if the MC estimates for the
marginal posterior are noisy and/or contain many discontinuities,
which is typically the case if the MC estimator is implemented using
particle filtering techniques.

Finally, the fourth category is transport map/dynamic rescaling methods
\citep[see e.g.][]{doi:10.1137/17M1134640,1903.03704}, which rely
on introducing a modified parameterization related to the original
parameterization via the nonlinear transport map. The transport map
is chosen so that the target distribution in the modified parameterization
is more well behaved and allows MCMC sampling using standard techniques.
The Dynamically rescaled Hamiltonian Monte Carlo (DRHMC) approach of \citet{1806.02068} involves a recipe for constructing transport maps
suitable for a large class of Bayesian hierarchical models, and where
the models are fitted using the (fixed scale) No-U-Turn Sampler (NUTS) Hamiltonian Monte
Carlo (HMC) algorithm \citep{Hoffman2014} implemented in Stan \citep{stan_user_guide}. 

The present paper also considers a transport map approach for Bayesian
hierarchical models, and sample from the modified target using HMC
methods. However, the strategy for constructing the transport map
considered here is different from that of DRHMC. Specifically,
DRHMC involves deriving the transport maps from the model specification itself,
and in particular it requires the availability of closed-form expressions for certain
precision- and Fisher information matrices associated with the model.
Moreover, the DRHMC approach is in practice limited to models containing only
a certain class of nonlinearities which lead to so-called constant
information parameterizations.   

Here, on the other hand, we consider transport maps derived from well-known importance
sampling (IS) methods for the latent variables only.
This approach relies only on the ability to evaluate the log-target density (and potentially it's derivatives)
pointwise,
and therefore bypasses the substantial analytic tractability requirement of DRHMC. 
The proposed approach is consequently more automatic in nature, and in particular
applicable to a wider range of nonlinear
models than DRHMC. Still, some analytical insight into the model 
is beneficial in terms of computational speed when choosing the initial iterates of
the involved iterative processes.

A fortunate property
of the proposed methodology, relative to conventional applications
of high-dimensional importance sampling \citep[see e.g.][]{KOOPMAN20092},
is that the importance densities applied within the present framework
may be of relatively low fidelity as long as they reflect the location
and scale of the distribution of the latent state conditioned both
on data and parameters. Since parameters and latent variables are
updated simultaneously, the slow exploration of the target associated
with Gibbs sampling is avoided. Moreover, being transport map-based,
rather than say RMHMC-based, the proposed methodology allows for the
application of standard HMC and in particular can be implemented with
minimal effort in Stan. 

The application of IS methods to construct transport maps also allows
the proposed methodology to be interpreted as a pseudo-marginal method,
namely a special case (with simulation sample size $n=1$) of the
pseudo-marginal HMC method of \citet{Lindsten2016}. However, our
focus on models with high-dimensional latent variables generally
precludes the application of `brute force' IS estimators that do not
reflect information from the data \citep[see, e.g.,][]{danielsson94}.
This is the case even for increased simulation sample size of the IS estimate,
as is possible in the general setup of \citet{Lindsten2016}. 

The rest of the paper is laid out as follows: Section~\ref{sec:Background}
provides some background and Section~\ref{sec:Transport-maps-based}
introduces IS-based transport maps. Section~\ref{sec:Specific-choices-of}
discusses specific choices of IS-based transport maps and Section
\ref{sec:Simulation-study} provides a simulation experiment where
the fidelity vs computational cost tradeoff of the different transport
maps is explored numerically. Finally, Section~\ref{sec:Realistic-application}
presents a realistic application and Section~\ref{sec:Discussion}
provides some discussion. The paper is accompanied by supplementary material
giving further details in several regards, and the code used for the computations
is available at \texttt{https://github.com/kjartako/TMHMC}.

\section{Background\label{sec:Background}}

This section outlines some background on HMC and why the application
of HMC in default formulations of hierarchical models is problematic.
In what follows, we use $\mathcal{N}(\mathbf{x}|{\boldsymbol{\mu}},{\boldsymbol{\Sigma}})$
to denote the probability density function of a $N(\boldsymbol{\mu},\boldsymbol{\Sigma})$
random vector evaluated at $\mathbf{x}$, while $\nabla_{\mathbf{z}}$
and $\nabla_{\mathbf{z}}^{2}$ are used, respectively, for the gradient/Jacobian
and Hessian operator with respect to the vector $\mathbf{z}$.

\subsection{HMC\label{sec:HMC}}

Over the past decade, HMC introduced by \citet{Duane1987} has been
extensively used as a general-purpose MCMC method, often applied for
simulating from posterior distributions arising in Bayesian models
\citep{neal2011mcmc}. HMC offers the advantage of producing close
to perfectly mixing MCMC chains by using the dynamics of a synthetic
Hamiltonian system as proposal mechanism. The popular
Bayesian modelling software Stan \citep{stan_user_guide} is an easy to use
HMC implementation based on the NUTS HMC algorithm of \citet{Hoffman2014}.

Suppose one seeks to sample from an analytically intractable target
distribution with density kernel $\tilde{\pi}(\mathbf{q}),\;\mathbf{q}\in\Omega\subseteq\mathbb{R}^{s}$.
To this end, HMC takes the variable of interest $\mathbf{q}$ as the
`position coordinate' of a Hamiltonian system, which is complemented
by an (artificial) `momentum variable' $\mathbf{p}\in\mathbb{R}^{s}$.
The corresponding Hamiltonian function specifying the total energy
of the dynamical system is given by 
\begin{equation}
H(\mathbf{q},\mathbf{p})=-\log\tilde{\pi}(\mathbf{q})+\frac{1}{2}\mathbf{p}'\mathbf{M}^{-1}\mathbf{p},\label{eq:Hamiltonian_initial}
\end{equation}
where $\mathcal{\mathbf{M}\in\mathbb{R}}^{s\times s}$ is a symmetric,
positive definite `mass matrix' representing an HMC tuning parameter.
For near-Gaussian target distributions, for instance, setting $\mathcal{\mathbf{M}}$
close to the precision matrix of the target ensures the best performance.
The law of motions under the dynamic system specified by the Hamiltonian
$H$ is determined by Hamilton's equations given by 
\begin{equation}
\frac{d}{dt}\mathbf{p}(t)=-\nabla_{\mathbf{q}}H\left(\mathbf{q}(t),\mathbf{p}(t)\right)=\nabla_{\mathbf{q}}\log\tilde{\pi}(\mathbf{q}),\qquad\frac{d}{dt}\mathbf{q}(t)=\nabla_{\mathbf{p}}H\left(\mathbf{q}(t),\mathbf{p}(t)\right)=\mathbf{M}^{-1}\mathbf{p}.\label{eq:Ham_eqns}
\end{equation}
It can be shown that the dynamics associated with Hamilton's equations
preserves both the Hamiltonian (i.e. $dH\left(\mathbf{q}(t),\mathbf{p}(t)\right)/dt=0)$
and the Boltzmann distribution $\pi(\mathbf{q},\mathbf{p})\propto\exp\{-H(\mathbf{q},\mathbf{p})\}\propto\tilde{\pi}(\mathbf{q})\;\mathcal{N}(\mathbf{p}|\mathbf{0}_{s},\mathbf{M})$,
in the sense that if $[\mathbf{q}(t),\mathbf{p}(t)]\sim\pi(\mathbf{q},\mathbf{p})$,
then $[\mathbf{q}(t+\tau),\mathbf{p}(t+\tau)]\sim\pi(\mathbf{q},\mathbf{p})$
for any (scalar) time increment $\tau$. Based on the latter property, a
valid MCMC scheme for generating $\{\mathbf{q}^{(k)}\}_{k}\sim\tilde{\pi}(\mathbf{q})$
would be to alternate between the following two steps: (i) Sample
a new momentum $\mathbf{p}^{(k)}\sim N(\mathbf{0}_{s},\mathbf{M})$
from the $\mathbf{p}$-marginal of the Boltzmann distribution; and
(ii) use the Hamiltonian's equations~(\ref{eq:Ham_eqns}) to propagate
$[\mathbf{q}(0),\mathbf{p}(0)]=[\mathbf{q}^{(k)},\mathbf{p}^{(k)}]$
for some increment $\tau$ to obtain $[\mathbf{q}(\tau),\mathbf{p}(\tau)]=[\mathbf{q}^{(k+1)},\mathbf{p}^{\ast}]$
and discard $\mathbf{p}^{\ast}$. However, for all but very simple
scenarios (like those with a Gaussian target $\tilde{\pi}(\mathbf{q})$)
the transition dynamics according to~(\ref{eq:Ham_eqns}) does not
admit closed-form solution, in which case it is necessary to rely
on numerical integrators for an approximative solution. Provided that
the numerical integrator used for that purpose is symplectic, the
numerical approximation error can be exactly corrected by introducing
an accept-reject (AR) step, which uses the Hamiltonian to compare
the total energy of the new proposal for the pair $(\mathbf{q},\mathbf{p})$
with that of the old pair inherited from the previous MCMC step \citep[see, e.g.,][]{neal2011mcmc}.
More specifically each iteration of the HMC algorithm involves the
following steps
\begin{itemize}
\item Refresh the momentum $\mathbf{p}^{(k)}\sim N(\mathbf{0}_{s},\mathbf{M})$.
\item Propagate approximately the dynamics~(\ref{eq:Ham_eqns}) from $(\mathbf{q}(0),\mathbf{p}(0))=(\mathbf{q}^{(k)},\mathbf{p}^{(k)})$
to obtain $(\mathbf{q}^{*},\mathbf{p}^{\text{*}})\approx(\mathbf{q}(L\varepsilon),\mathbf{p}(L\varepsilon))$
using $L$ symplectic integrator steps with time-step size $\varepsilon$.
\item Set $\mathbf{q}^{(k+1)}=\mathbf{q}^{*}$ with probability $\min(1,\exp(H(\mathbf{q}^{(k)},\mathbf{p}^{(k)})-H(\mathbf{q}^{*},\mathbf{p}^{*}))$
and $\mathbf{q}^{(k+1)}=\mathbf{q}^{(k)}$ with remaining probability.
\end{itemize}
The most commonly used symplectic integrator is the Störmer-Verlet
or leapfrog integrator \citep[see, e.g.,][]{Leimkuhler:2004,neal2011mcmc}.
When implementing numerical integrators with AR-corrections it is
critical that the selection of the step size accounts for the inherent
trade-off between the computing time required for generating AR proposals
and their quality reflected by their corresponding acceptance rates.
$(\mathbf{q},\mathbf{p})$-proposals generated by using small (big)
step sizes tend to be computationally expensive (cheap) but imply
a high (low) level of energy preservation and thus high (low) acceptance
rates. Finally, the energy preservation properties of the symplectic
integrator for any given step size critically relies on the nature
of the target distribution. It is taken as a rule of
thumb for the remainder of the text that high-dimensional, highly
non-Gaussian targets typically require small step sizes and many steps,
whereas high-dimensional near-Gaussian targets can be sampled efficiently
with rather large step sizes and few steps.

\subsection{Hierarchical models and HMC\label{sec:PM_HMC}}

Consider a stochastic model for a collection of observed data $\mathbf{y}$
involving a collection of latent variables $\mathbf{x}$ and a vector
of parameters $\boldsymbol{\theta}\in\mathbb{R}^{d}$ with prior density
$p(\boldsymbol{\theta})$. The conditional likelihood for observations
$\mathbf{y}$ given a value of the latent variable $\mathbf{x}\in\mathbb{R}^{D}$
is denoted by $p(\mathbf{y}|\mathbf{x},\boldsymbol{\theta})$ and
the prior for $\mathbf{x}$ by $p(\mathbf{x}|\boldsymbol{\theta})$.
This latent variable model is assumed to be nonlinear and/or non-Gaussian
so that both the joint posterior for $(\mathbf{x},\boldsymbol{\theta})$
as well as the marginal posterior for $\boldsymbol{\theta}$ are analytically
intractable.

The joint posterior for $(\mathbf{x},\boldsymbol{\theta})$ under
such a latent variable model, given by $p(\mathbf{x},\boldsymbol{\theta}|\mathbf{y})\propto p(\mathbf{y}|\mathbf{x},\boldsymbol{\theta})p(\mathbf{x}|\boldsymbol{\theta})p(\boldsymbol{\theta})$,
can have a complex dependence structure. In particular, when the scale
of $\mathbf{x}|\boldsymbol{\theta},\mathbf{y}$ varies substantially
as a function of $\boldsymbol{\theta}$ in the typical range of $p(\boldsymbol{\theta}|\mathbf{y})$,
the joint posterior will be ``funnel-shaped'' \citep[see][Figure 1 for an illustration]{1806.02068}.
In this case, the HMC algorithm, as described in Section~\ref{sec:HMC},
for $\mathbf{q}=(\mathbf{x}^{T},\boldsymbol{\theta}^{T})^{T}$ must
be tuned for the most extremely scaled parts of the target distribution
to ensure exploration of the complete target distribution. This, in
turn lead to a computationally wasteful exploration of the more moderately
scaled parts of the target, as the tuning parameters cannot themselves
depend on $\mathbf{q}$ (under regular HMC). In addition, automated
tuning of integrator step sizes (and mass matrices) crucially relies on the most extremely scaled parts being visited during the initial tuning phase. If not, they may not be explored at all.

\section{Transport maps based on IS densities\label{sec:Transport-maps-based}}

To counteract such undesired extreme tuning, while avoiding computationally
costly $\mathbf{q}$-dependent tuning such as RMHMC, the approach
taken here involves ``preconditioning'' the original target so that
the resulting modified target is close to Gaussian and thus suitable
for statically tuned HMC. Such preconditioning with the aim of producing
more tractable target distributions for MCMC methods have a long tradition,
and prominent examples are the affine re-parameterizations common
for Gibbs sampling applied to regression models \citep[see, e.g.,][Chapter 12]{GelmanBDA3}.
More recent approaches with such ends involve semi-parametric transport
map approach of \citet{doi:10.1137/17M1134640}, and, neural transport
as described by \citet{1903.03704}. The approach taken here share
many similarities with the dynamically rescaled HMC approach of \citet{1806.02068},
but the strategy for constructing the transport map considered here
is very different and is applicable to more general
models. 

In a nutshell, a transport map, say $T$, is a smooth bijective mapping relating the
original parameterization $\mathbf q \sim \pi_{\mathbf q}(\mathbf q)$ and some modified 
parameterization $\mathbf q^\prime$
via $\mathbf q = T(\mathbf q^\prime)$. If $\mathbf q^\prime $ is some random draw 
$\sim \pi_{\mathbf q^\prime}(\mathbf q^\prime) = 
\pi_{\mathbf q}(T(\mathbf q^\prime))|\nabla_{\mathbf q^\prime} T(\mathbf q^\prime)|$,
then a draw distributed according to $\pi_{\mathbf q}$ is achieved by simply applying the transport
map to $\mathbf q^\prime$. The aim of introducing this construction, is that $T$ can be
chosen so that $\pi_{\mathbf q^\prime}$ is loosely speaking "more suitable for MCMC sampling". 
In practice, this rather vague aim is replaced by
making $\pi_{\mathbf q^\prime}$ close to a Gaussian distribution with independent components,
which can be sampled very efficiently using HMC.

\subsection{Transport maps for Bayesian hierarchical models}

In the current situation involving a Bayesian hierarchical
model, a transport map $T$ that is non-trivial for the latent variables only, 
\[
\mathbf q =\left[\begin{array}{c} \boldsymbol \theta \\ \mathbf x \end{array} \right]
= T(\mathbf q^\prime) = 
\left[\begin{array}{c} \boldsymbol \theta \\ \gamma_{\boldsymbol{\theta}}(\mathbf{u}) \end{array} \right],
\; \mathbf q^\prime = \left[\begin{array}{c}\boldsymbol \theta \\ \mathbf u\end{array}\right],
\]
is considered.
The transport map specific to the latent variables, $\gamma_{\boldsymbol{\theta}}:\mathbb R^D \rightarrow \mathbb R^D$ is assumed
to be a smooth
bijective mapping
for each $\boldsymbol{\theta}$. 
As we have $\nabla_{\mathbf u} \boldsymbol \theta = \mathbf 0$ in the above transport map, it follows that
$|\nabla_{\mathbf q^\prime} T(\mathbf q^\prime)|=|\nabla_{\mathbf{u}}\gamma_{\boldsymbol{\theta}}(\mathbf{u})|$, and
thus the modified target distribution has the form:
\begin{equation}
\tilde{\pi}(\boldsymbol{\theta},\mathbf{u}|\mathbf{y})\propto
|\nabla_{\mathbf{u}}\gamma_{\boldsymbol{\theta}}(\mathbf{u})|p(\boldsymbol{\theta})
\left[p(\mathbf{y}|\mathbf{x},\boldsymbol{\theta})p(\mathbf{x}|\boldsymbol{\theta})\right]_{\mathbf{x}=\gamma_{\boldsymbol{\theta}}(\mathbf{u})}.\label{eq:t-target}
\end{equation}
Notice in particular that the original
parameterization of the latent variables is computed in each evaluation
of~(\ref{eq:t-target}), and thus obtaining MCMC samples in the 
$(\boldsymbol{\theta},\mathbf{x})=(\boldsymbol{\theta},\gamma_{\boldsymbol{\theta}}(\mathbf{u}))$
parameterization comes at no additional cost when MCMC samples targeting~
(\ref{eq:t-target}) are available.

Further, let $m(\mathbf{x}|\boldsymbol{\theta})$ denote the density
of $\gamma_{\boldsymbol{\theta}}(\mathbf{u})$ \textit{when} $\mathbf{u}\sim N(\mathbf{0}_{D},\mathbf{I}_{D})$.
In particular, $m(\mathbf{x}|\boldsymbol{\theta})$ is implicitly
related to the underlying standard Gaussian distribution via the change
of variable formula: $\mathcal{N}(\mathbf{u}|\mathbf{0}_{D},\mathbf{I}_{D})=|\nabla_{\mathbf{u}}\gamma_{\boldsymbol{\theta}}(\mathbf{u})|\left[m(\mathbf{x}|\boldsymbol{\theta})\right]_{\mathbf{x}=\gamma_{\boldsymbol{\theta}}(\mathbf{u})}$.
Consequently, eliminating the Jacobian determinant in~(\ref{eq:t-target})
results in 
\begin{align}
\tilde{\pi}(\boldsymbol{\theta},\mathbf{u}|\mathbf{y})\propto\mathcal{N}(\mathbf{u}|\mathbf{0}_{D},\mathbf{I}_{D})p(\boldsymbol{\theta})\omega_{\boldsymbol{\theta}}(\mathbf{u}), & \;\omega_{\boldsymbol{\theta}}(\mathbf{u})=\left[\frac{p(\mathbf{y}|\mathbf{x},\boldsymbol{\theta})p(\mathbf{x}|\boldsymbol{\theta})}{m(\mathbf{x}|\boldsymbol{\theta})}\right]_{\mathbf{x}=\gamma_{\boldsymbol{\theta}}(\mathbf{u})}.\label{eq:t-target-2}
\end{align}
Representation~(\ref{eq:t-target-2}) reveal that if $m(\mathbf{x}|\boldsymbol{\theta})=p(\mathbf{x}|\mathbf{y},\boldsymbol{\theta})$
(i.e. $\gamma_{\boldsymbol{\theta}}(\mathbf{u})\sim\mathbf{x}|\mathbf{y},\boldsymbol{\theta}$),
the parameters and latent variables exactly ``decouples'' and~(\ref{eq:t-target})
and~(\ref{eq:t-target-2}) reduces to $\mathcal{N}(\mathbf{u}|\mathbf{0}_{D},\mathbf{I}_{D})p(\boldsymbol{\theta}|\mathbf{y})$
\citep[see also][for a similar discussion]{Lindsten2016}. Such a
situation will be well suited for HMC sampling (provided of course
that the marginal likelihood $p(\boldsymbol{\theta}|\mathbf{y})$
is reasonably well-behaved). Of course, such an ideal situation is
in practice unattainable when the model in question is nonlinear/non-Gaussian
as neither $p(\boldsymbol{\theta}|\mathbf{y})$ nor $p(\mathbf{x}|\mathbf{y},\boldsymbol{\theta})$
will have analytical forms. The strategy pursued here is therefore
to take $m(\mathbf{x}|\boldsymbol{\theta})$ as an approximation to
$p(\mathbf{x}|\mathbf{y},\boldsymbol{\theta})$ in order to obtain
an approximate decoupling effect, i.e. so that $\omega_{\boldsymbol{\theta}}(\mathbf{u})$
is fairly flat across the region where $\mathcal{N}(\mathbf{u}|\mathbf{0}_{D},\mathbf{I}_{D})$
has significant probability mass.

\subsection{Relation to importance sampling and pseudo-marginal methods\label{subsec:Relation-to-importance}}

The $\omega_{\boldsymbol{\theta}}(\mathbf{u})$ of~(\ref{eq:t-target-2})
is recognized to be an importance weight targeting the marginal likelihood
$p(\mathbf{y}|\boldsymbol{\theta})$ (i.e. $E_{\mathbf u}(\omega_{\boldsymbol{\theta}}(\mathbf{u}))=p(\mathbf{y}|\boldsymbol{\theta})$) when $\mathbf{u}\sim N(\mathbf{0}_{D},\mathbf{I}_{D})$.
This observation is important for at least three reasons. Firstly,
it is clear that the large literature on importance sampling- and
similar methods for hierarchical models \citep[among many others,][]{Shephard1997,Richard2007,RSSB:RSSB700,durbin_koopman_2ed}
may be leveraged to suggest suitable choices for importance density
$m(\mathbf{x}|\boldsymbol{\theta})$ or $\gamma_{\boldsymbol{\theta}}(\mathbf{u})$.
Specific choices considered here are discussed in more detail in Section~
\ref{sec:Specific-choices-of}.

Secondly, as discussed, e.g., in \citet{KOOPMAN20092}, importance
sampling-based likelihood estimates such as $\omega_{\boldsymbol{\theta}}(\mathbf{u})$
may have infinite variance and thus become unreliable, in particular
in high-dimensional applications. This occurs when the tails of $m(\mathbf{x}|\boldsymbol{\theta})$
are thinner than those of the target distribution $p(\mathbf{x}|\boldsymbol{\theta},\mathbf{y})\propto p(\mathbf{y}|\mathbf{x},\boldsymbol{\theta})p(\mathbf{x}|\boldsymbol{\theta})$,
making $\omega_{\boldsymbol{\theta}}(\mathbf{u})$ unbounded as a
function of $\mathbf{u}$. However, under the modified target~(\ref{eq:t-target-2})
the likelihood estimate is combined with the thin-tailed standard
normal distribution in $\mathbf{u}$, which counteracts the potential
unboundedness of the IS weight in the $\mathbf{u}$-direction. This
robustness with respect to the infinite-variance problem is also evident
in the representation~(\ref{eq:t-target}) of the target, which does
not explicitly involve the importance sampling weight. Affine transport
maps $\gamma_{\boldsymbol{\theta}}(\mathbf{u})$, and consequently
thin-tailed Gaussian importance densities $m(\mathbf{x}|\boldsymbol{\theta})$,
lead to the Jacobian determinant $|\nabla_{\mathbf{u}}\gamma_{\boldsymbol{\theta}}(\mathbf{u})|$
being constant with respect to $\mathbf{u}$. Consequently, in this case the tail
behavior of~(\ref{eq:t-target}) with respect to $\mathbf{u}$ will
be the same as the tail behavior of $p(\boldsymbol{\theta},\mathbf{x}|\mathbf{y})$
in $\mathbf{x}$. Thus, the proposed methodology may be seen as a resolution
of the infinite variance problems complicating the application of high-dimensional
importance sampling.

Finally, the proposed methodology may be seen as a special
case of the pseduo-marginal HMC (PM-HMC) method of \citet{Lindsten2016}.
PM-HMC relies on joint HMC sampling of a Monte Carlo estimate of the
marginal likelihood and the random variables used to generate said
estimate. \citet{Lindsten2016} find a similar decoupling effect
by admitting their Monte Carlo estimate be based on $n\geq1$ importance
weights (at the cost of increasing the dimensionality of $\mathbf{u}$
in their counterpart to~(\ref{eq:t-target-2})), and are to a lesser
degree reliant on choosing high-quality importance densities. In particular,
\citet{Lindsten2016} use $m(\mathbf{x}|\boldsymbol{\theta})=p(\mathbf{x}|\boldsymbol{\theta})$
in their illustrations, which for moderately dimensional and low-signal-to
noise situations will produce a good decoupling effect for moderate
$n$. However, in the present work we focus on high-dimensional applications
where it is well known that such ``brute force'' importance sampling
estimators can suffer from prohibitively large variances for any practical
$n$ \citep[see, e.g.,][]{danielsson94}, and thus focus rather on
higher fidelity importance densities and $n=1$. 

\citet{Lindsten2016} also propose a symplectic integrator suitable
for HMC applications with target distributions on the form~(\ref{eq:t-target-2})
under the ``close to decoupling'' assumption. In
the decoupling case $\mathbf{u}\mapsto\omega_{\boldsymbol{\theta}}(\mathbf{u})\propto1$,
the integrator reduces to a standard leapfrog integrator in the dynamics
of $\boldsymbol{\theta}$, whereas the dynamics of $\mathbf{u}$ (typically
high-dimensional) are simulated exactly. This integrator
will be referred to as the LD-integrator in the example applications
and is detailed in the supplementary material, Section~\ref{sec:The--integrator}.

\section{Specific choices of $m(\mathbf{x}|\boldsymbol{\theta})$ and $\gamma_{\boldsymbol{\theta}}(\mathbf{u})$\label{sec:Specific-choices-of}}

As alluded to above, taking $m(\mathbf{x}|\boldsymbol{\theta})=p(\mathbf{x}|\boldsymbol{\theta})$
may in cases where data $\mathbf{y}$ are rather un-informative with
respect to the latent variable $\mathbf{x}$ lead to satisfactory
results \citep[see e.g.][Section 2.5]{stan_user_guide}. However,
as illustrated by e.g. \citet{1806.02068}, such procedures can lead
to misleading MCMC results if data are more informative with respect
to the latent variables. 
An even more challenging situation with 
$m(\mathbf{x}|\boldsymbol{\theta})=p(\mathbf{x}|\boldsymbol{\theta})$
is when one or more elements of $\boldsymbol{\theta}$ determine how
informative the data are with respect to the latent variables (e.g.
$\sigma$ when $y_{i}\sim N(x_{i},\sigma^{2})$), as this may still lead
to a funnel-shaped target distribution. On the other hand, as illustrated
by \citet{1806.02068}, rather crude transport maps reflecting only
roughly the location and scale of $p(\mathbf{x}|\mathbf{y},\boldsymbol{\theta})$
may lead to dramatic speedups, and the resolution of funnel-related
problems. In the rest of this section, two families of strategies
for locating transport maps are discussed. Both are well
known in the context of importance sampling, and are
typically applicable when $p(\mathbf{x}|\boldsymbol{\theta})$ is
non-Gaussian. 

\subsection{$m(\mathbf{x}|\boldsymbol{\theta})$ and $\gamma_{\boldsymbol{\theta}}(\mathbf{u})$
derived from approximate Laplace approximations}

As explained e.g. in \citet{RSSB:RSSB700}, the Laplace approximation
(also often referred to as the second order approximation) for integrating
out latent variables relies on approximating $p(\mathbf{x}|\mathbf{y},\boldsymbol{\theta})$
with a $N(\mathbf{h}_{\boldsymbol{\theta}},\mathbf{G}_{\boldsymbol{\theta}}^{-1})$
density, where
\begin{align*}
\mathbf{h}_{\boldsymbol{\theta}} & =\arg\max_{\mathbf{x}}\log\left[p(\mathbf{x}|\boldsymbol{\theta})p(\mathbf{y}|\mathbf{x},\boldsymbol{\theta})\right],\\
\mathbf{G}_{\boldsymbol{\theta}} & =-\nabla_{\mathbf{x}}^{2}\log\left[p(\mathbf{x}|\boldsymbol{\theta})p(\mathbf{y}|\mathbf{x},\boldsymbol{\theta})\right]_{\mathbf{x}=\mathbf{h}_{\boldsymbol{\theta}}}.
\end{align*}
Namely, the first and second order derivatives of $\log p(\mathbf{x}|\mathbf{y},\boldsymbol{\theta})$
at the mode are matched with the same derivatives of the approximating
Gaussian log-density. Due to conditional independence assumptions
often involved in modelling, the negative Hessian of $-\log p(\mathbf{x}|\mathbf{y},\boldsymbol{\theta})$
is typically sparse which, when exploited, can substantially speed
up the associated Cholesky factorizations. 

In the present situation, obtaining the exact mode $\mathbf{h}_{\boldsymbol{\theta}}$ is typically not
desirable from a computational perspective. Rather, given an initial
guesses for $\mathbf{h}_{\boldsymbol{\theta}}$ and $\mathbf{G}_{\boldsymbol{\theta}}$,
say $\mathbf{h}_{\boldsymbol{\theta}}^{(0)}$ and $\mathbf{G}_{\boldsymbol{\theta}}^{(0)}$,
a sequence of gradually more refined approximate solutions 
$\mathbf{h}_{\boldsymbol{\theta}}^{(k)}$ and $\mathbf{G}_{\boldsymbol{\theta}}^{(k)}$ are calculated
via iterations of Newton's method for optimization or an approximation thereof (see supplementary material,
Sections~\ref{sec:Details-related-to-sim-study} and~\ref{sec:Details-related-to} for details specific to the
models considered shortly).

Finally, for some fixed number of iterations, $K=0,1,2,\dots$, the
transport map is taken to be
\begin{equation}
\gamma_{\boldsymbol{\theta}}(\mathbf{u})=\mathbf{h}_{\boldsymbol{\theta}}^{(K)}+\left(\mathbf{L}_{\boldsymbol{\theta}}^{(K)}\right)^{-T}\mathbf{u},\label{eq:laplace_transport}
\end{equation}
where $\mathbf{L}^{(K)}$ is the lower triangular Cholesky factor
of $\mathbf{G}_{\boldsymbol{\theta}}^{(K)}$, so that $m(\mathbf{x}|\boldsymbol{\theta})=\mathcal{N}\left(\mathbf{x}|\mathbf{h}_{\boldsymbol{\theta}}^{(K)},\left[\mathbf{G}_{\boldsymbol{\theta}}^{(K)}\right]^{-1}\right)$.
Notice in particular that the Jacobian determinant of $\gamma_{\boldsymbol{\theta}}(\mathbf{u})$,
required in representation~(\ref{eq:t-target}) (or in the normalization
constant of $m(\mathbf{x}|\boldsymbol{\theta})$ in~(\ref{eq:t-target-2})),
takes a particularly simple form, namely $|\nabla_{\mathbf{u}}\gamma_{\boldsymbol{\theta}}(\mathbf{u})|=|\mathbf{L}_{\boldsymbol{\theta}}^{(K)}|^{-1}$,
when applying the affine transport map~(\ref{eq:laplace_transport}).
 It should be noted
that the applicability of the Laplace approximation relies critically
on that $p(\mathbf{x}|\mathbf{y},\boldsymbol{\theta})$ is unimodal
and log-concave in a region around the mode that also contains $\mathbf{h}_{\boldsymbol{\theta}}^{(0)}$.

Choices of $\mathbf{h}_{\boldsymbol{\theta}}^{(0)}$, $\mathbf{G}_{\boldsymbol{\theta}}^{(0)}$ 
and the iteration over $k$ are inherently
model specific. However, for a rather general class of models,
the initial guesses may be taken to be 
\begin{eqnarray}
\mathbf{G}_{\boldsymbol{\theta}}^{(0)}&=&\mathbf{G_{\boldsymbol{\theta},x}}+\mathbf{G_{\boldsymbol{\theta},y|\mathbf{x}}}
\label{eq:laplace_guess_G}\\
\mathbf{h}_{\boldsymbol{\theta}}^{(0)}&=&\left(\mathbf{G}_{\boldsymbol{\theta}}^{(0)}\right)^{-1}
(\mathbf{G_{\boldsymbol{\theta},\mathbf{x}}}\mathbf{h}_{\boldsymbol{\theta},\mathbf{x}}+
\mathbf{G_{\boldsymbol{\theta},y|\mathbf{x}}h_{\boldsymbol{\theta},y|\mathbf{x}}}),\label{eq:laplace_guess}
\end{eqnarray}
where $\mathbf{h}_{\boldsymbol{\theta},\mathbf{x}}$ and $\mathbf{G_{\boldsymbol{\theta},x}}$
are the mean and precision matrix associated with $\mathbf x | \boldsymbol \theta$. 
Further,
$\mathbf{h}_{\boldsymbol{\theta},\mathbf{y}|\mathbf{x}}$ and $\mathbf{G_{\boldsymbol{\theta},y|\mathbf{x}}}$
are the mode, and the negative Hessian at the mode
of $\mathbf{x}\mapsto\log p(\mathbf{y}|\mathbf{x},\boldsymbol{\theta})$.
Note that Equations~\ref{eq:laplace_guess_G} and~\ref{eq:laplace_guess}
correspond to the precision and mean of the crude approximation $\propto\mathcal{N}\left(\mathbf{x}|\mathbf{h}_{\boldsymbol{\theta},\mathbf{x}},\mathbf{G_{\boldsymbol{\theta},\mathbf{x}}}^{-1}\right)\mathcal{N}\left(\mathbf{x}|\mathbf{h}_{\boldsymbol{\theta},\mathbf{y}|\mathbf{x}},\mathbf{G_{\boldsymbol{\theta},y|\mathbf{x}}}^{-1}\right)$
to $p(\mathbf{x}|\boldsymbol{\theta})p(\mathbf{y}|\mathbf{x},\boldsymbol{\theta})$.
Moreover, it is also in some cases possible to find approximations
to the involved negative Hessian that do not depend on $\mathbf{x}$
\citep[see e.g.][]{1806.02068}, reducing the number of Cholesky factorization
per evaluation of~(\ref{eq:t-target}) to one.

Interestingly, the approximate pseudo-marginal MCMC method of \citet{Gomez-Rubio2018}
is closely connected to the proposed methodology with Laplace approximation-based
transport maps. Specifically, $\omega_{\boldsymbol{\theta}}(\mathbf{0}_{D})$
is the conventional Laplace approximation \citep[see e.g.][]{doi:10.1080/01621459.1986.10478240}
of $p(\mathbf{y}|\boldsymbol{\theta})$ (modulus the usage of an approximate
mode and Hessian).
By substituting $\omega_{\boldsymbol{\theta}}(\mathbf{0}_{D})$ for
$\omega_{\boldsymbol{\theta}}(\mathbf{u})$ in~(\ref{eq:t-target})
(and integrating analytically over $\mathbf{u}$), the target distribution
of \citet{Gomez-Rubio2018} is obtained. Thus, the proposed methodology
with Laplace approximation-based transport maps may be regarded as
variant of the \citet{Gomez-Rubio2018} method that corrects for the
approximation error of the underlying Laplace approximation. 

\subsection{$m(\mathbf{x}|\boldsymbol{\theta})$ and $\gamma_{\boldsymbol{\theta}}(\mathbf{u})$
derived from the Efficient Importance Sampler}

The efficient importance sampler (EIS) algorithm of \citet{Richard2007}
is a widely used technique for constructing close to optimal importance
densities, typically in the context of integrating out latent variables.
At its core, the EIS relies initially on eliciting a family of sampling
mechanisms, say $\mathbf{x}=\Gamma_{\mathbf{a}}(\mathbf{u})$, $\Gamma_{\mathbf{a}}:\mathbb{R}^{D}\mapsto\mathbb{R}^{D}$,
indexed by some, typically high-dimensional parameter $\mathbf{a}\in\mathcal{A}$.
Moreover, for all $\mathbf{a}\in\mathcal{A}$, and for $\mathbf{u}\sim N(\mathbf{0}_{D},\mathbf{I}_{D})$,
the density of $\Gamma_{\mathbf{a}}(\mathbf{u})$ is denoted by $m_{\mathbf{a}}(\mathbf{x})$.
The EIS algorithm proceeds by first sampling a collection of ``common
random numbers'' $\mathbf{Z}=\left\{ \mathbf{z}^{(i)}\right\}_{i=1}^r ,\;\mathbf{z}^{(i)}\sim\text{iid}\;N(\mathbf{0}_{D},\mathbf{I}_{D}),\;i=1,\dots,r$,
then selecting an initial parameter $\mathbf{a}^{[0]}$, and finally
iterate over the below steps for $j=1,\dots,J$:
\begin{itemize}
\item Sample latent states $\mathbf{x}^{(i)}=\Gamma_{\mathbf{a}^{[j-1]}}(\mathbf{z}^{(i)}),\;i=1,\dots,r$. 
\item Locate a new $\mathbf{a}^{[j]}$ as a (generally approximate) minimizer
(over $\mathbf{a}$) of the sample variance of the importance weights
$w_{\mathbf{a}}^{(i)}=p(\mathbf{y}|\mathbf{x}^{(i)},\boldsymbol{\theta})p(\mathbf{x}^{(i)}|\boldsymbol{\theta})/m_{\mathbf{a}}(\mathbf{x}^{(i)}),\;i=1,\dots,r$. 
\end{itemize}
An unbiased estimate of $p(\mathbf{y}|\boldsymbol{\theta})$ is given
by the means of conventional importance sampling \citep[Section 3.3]{Robert2004}
based on importance density $m_{\mathbf{a}^{[J]}}(\mathbf{x})$, with
random draws (from $m_{\mathbf{a}^{[J]}}(\mathbf{x})$) generated
based on random numbers independent from $\mathbf{z}^{(i)},\;i=1,\dots,n$. 

Notice that the near optimal EIS parameter $\mathbf{a}^{[J]}=\mathbf{a}^{[J]}(\boldsymbol{\theta},\mathbf{Z})$
generally depends both on $\boldsymbol{\theta}$ and $\mathbf{Z}$.
In the present context, for some fixed set of common random numbers
$\mathbf{Z}$ and number of EIS iterations $J$, the importance density
of~(\ref{eq:t-target-2}) is simply set equal to the EIS importance
density, i.e. $m(\mathbf{x}|\boldsymbol{\theta})=m_{\mathbf{a}^{[J]}(\boldsymbol{\theta},\mathbf{Z})}(\mathbf{x})$.
Notice in particular that the EIS iterations above must be repeated
for each evaluation of~(\ref{eq:t-target-2}), and that the common
random numbers must be kept fixed during each HMC iteration (which
typically involve several evaluations of~(\ref{eq:t-target-2}) and
its gradient), or throughout the whole MCMC simulation.

The EIS importance density is often regarded as more reliable than
the Laplace approximation counterpart, as it explicitly seeks to minimize
the importance weight variation across typical outcomes of importance
density. In addition, the family of importance densities $m_{\mathbf{a}}(\mathbf{x})$
may be constructed to highly non-Gaussian densities, whereas the Laplace
approximation importance density is multivariate Gaussian. On the
other hand, the EIS algorithm typically is substantially more costly
in a computational perspective, whether this additional computational
effort pays of in terms of a better decoupling effect in~(\ref{eq:t-target},\ref{eq:t-target-2})
is sought to be answered here.

The sketch of the EIS algorithm above is intentionally kept somewhat
vague, as the actual details, both in terms of selecting $m_{\mathbf{a}}(\mathbf{x})$
and how the optimization step is implemented, depends very much on
the model specification at hand. 
A more detailed description of the EIS suitable for the models considered
in the simulation study discussed shortly is given in Section~\ref{sec:The-EIS-principle}
of the supplementary material. 

\subsection{Implementation and Tuning Parameters\label{subsec:Implementation-and-Tuning}}

The proposed methodology has been implemented in two ways. Firstly,
the Laplace approximation-based methods are implemented in Stan using the modified target
representation~(\ref{eq:t-target}). This is also the case for the reference
method corresponding to $m_{\boldsymbol{\theta}}(\mathbf{x})=p(\mathbf{x}|\boldsymbol{\theta})$.

Secondly, we also consider a bespoke HMC implementation as outlined in Section~\ref{sec:HMC},
for $\mathbf{q}=(\boldsymbol{\theta}^{T},\mathbf{u}^{T})^{T}$, targeting
either~(\ref{eq:t-target}, for Laplace approximation-based methods)
or~(\ref{eq:t-target-2}, for EIS-based methods).
This HMC method is based on the LD-integrator (see supplementary material,
Section~\ref{sec:The--integrator}) in order to better exploit the
approximate decoupling effects in the target, and was in
particular included to explore the advantage of using the LD-integrator
over the leapfrog integrator in the present situation. 

The mass matrix in the bespoke implementation was taken to be 
\[
\mathbf{M}=\left[\begin{array}{cc}
\hat{\mathbf{M}}_{\boldsymbol{\theta}} & \mathbf{0}_{d\times D}\\
\mathbf{0}_{D\times d} & \mathbf{I}_{D}
\end{array}\right],
\]
where $\hat{\mathbf{M}}_{\boldsymbol{\theta}}=-\nabla_{\boldsymbol{\theta}}^{2}\log\left[\hat{p}(\mathbf{y}|\boldsymbol{\theta})p(\boldsymbol{\theta})\right]_{\boldsymbol{\theta}=\hat{\boldsymbol{\theta}}}$
and the simulated MAP $\hat{\boldsymbol{\theta}}=\arg\max_{\boldsymbol{\theta}}\log\left[\hat{p}(\mathbf{y}|\boldsymbol{\theta})p(\boldsymbol{\theta})\right]$
is obtained from an EIS importance sampling estimate $\hat{p}(\mathbf{y}|\boldsymbol{\theta})$
of $p(\mathbf{y}|\boldsymbol{\theta})$. Finding the approximate parameter
marginal posterior precision $\hat{\mathbf{M}}_{\boldsymbol{\theta}}$
is very fast and requires minimal additional effort as gradients of
the importance weight with respect to $\boldsymbol{\theta}$ are already
available via automatic differentiation (AD, to be discussed shortly).
Notice that the mass matrix specific to $\mathbf{u}$ is take to be
the identity to match the precision of the $N(\mathbf{0}_{D},\mathbf{I}_{D})$
``prior'' of $\mathbf{u}$ in~(\ref{eq:t-target},\ref{eq:t-target-2}).
As for the integrator step size $\varepsilon$ and the number of integrator
steps $L$, we retain $L$ as a tuning parameter while keeping the
total integration time $\varepsilon L$ per HMC proposal fixed at
$\approx\pi/2$. This choice of total integration time is informed
by the expectation that $\boldsymbol{\theta},\mathbf{u}|\mathbf{y}$
under~(\ref{eq:t-target},\ref{eq:t-target-2}) will be close to a
Gaussian with precision matrix $\mathbf{M}$. Moreover, whenever $\tilde{\pi}(\mathbf{q})$
in~(\ref{eq:Hamiltonian_initial}) is Gaussian with precision $\mathbf{M}$,
the dynamics~(\ref{eq:Ham_eqns}) are periodic with period $t=2\pi$,
and choosing a quarter of such a cycle leads to HMC proposals $\mathbf{q}^{*}$
independent of the current configuration $\mathbf{q}^{(k)}$ \citep[see e.g.][]{neal2011mcmc,mannseth2018}.
Finally, $L$ is tuned by hand to obtain acceptance rates around $0.9$.

Both implementations rely on the ability to compute gradients of log-targets~
(\ref{eq:t-target},\ref{eq:t-target-2}) with respect to both $\boldsymbol{\theta}$
and $\mathbf{u}$. To this end, we rely on Automatic Differentiation
(AD). In Stan, this is done automatically, whereas in the bespoke
implementation, the Adept C++ automatic differentiation software library
\citep{Hogan2014} is applied. Notice that for the Laplace approximation-based method, AD is applied
to calculations of band-Cholesky factorizations, and thus there may be room for improvement
in CPU times if the AD libraries supported such operations natively.
The bespoke algorithm is implemented using the $\textsf{R}$ \citep{R} package
\texttt{\textbf{Rcpp}} by \citet{Rcpp}, which makes it possible to
run compiled C++ code in $\textsf{R}$. Stan is used through
its $\textsf{R}$ interface \texttt{\textbf{rstan}} \citep{Rstan}, version 2.19.2.
The same C++ compiler was used for both the bespoke and Stan methods.
All computations are performed using $\textsf{R}$ version 3.6.1 on
a PC with an Intel Core i5-6500 processor running at 3.20 GHz. 

\section{Simulation study\label{sec:Simulation-study}}

This section presents applications of the proposed methodology to
three non-Gaussian/nonlinear state-space latent variable models for
the purpose of benchmarking against alternative methods. State-space models with univariate state were chosen as the Laplace approximation-based
methods only require tri-diagonal Cholesky factorizations, which are
easily implemented in the Stan language. The specific models are selected
to illustrate the performance under different, empirically relevant,
scenarios. In particular, the three models exhibit significantly different,
and variable signal-to-noise ratios, which as discussed above may
modulate the need for (non-trivial) transport map methods. 

In the proceeding, different combinations of implementation ($\in\{$Stan,
LD$\}$) and transport map method ($\in\{$Prior, Laplace, EIS, Fisher$\}$)
are considered, where ``LD'' refers to the bespoke HMC implementation
with LD integrator. Transport map ``Prior'' correspond to $m_{\boldsymbol{\theta}}(\mathbf{x})=p(\mathbf{x}|\boldsymbol{\theta})$
and is equivalent to carrying out the simulations in an $(\boldsymbol{\theta},\boldsymbol{\eta})$-parameterization
where $\boldsymbol{\eta}=(\eta_{1},\dots,\eta_{D})^{\prime}$ are
a-priori standard normal disturbances of the models to be discussed. 

Transport map method ``Fisher'' corresponds to Fisher information-based DRHMC
approach of \citet{1806.02068} applied to the latent variables only (i.e. general DRHMC involves non-trivial transport
maps for the parameters also).
Fisher also leads to an affine transport map $\gamma_{\boldsymbol{\theta}}(\mathbf{u})=\mathbf{h}_{F}+\mathbf{L}_{F}^{-T}\mathbf{u}$,
$\mathbf{L}_{F}\mathbf{L}_{F}^{T}=\mathbf{G}_{F}$. Here, $\mathbf{G}_{F}$
is the sum of the a-priori precision matrix of $\mathbf{x}$ and the
Fisher information of the observations with respect to $\mathbf{x}$.
Notice that this method requires
both that said Fisher information is constant with respect to the
latent state, and that the $p(\mathbf x |\boldsymbol \theta)$ precision matrix has closed
form, where the latter requirement limits its applicability to the
first two models considered below.

Methods LD-Prior and LD-Fisher were not carried out as the default
tuning discussed in Section~\ref{subsec:Implementation-and-Tuning}
work poorly in these cases. Moreover, Stan-EIS was also not considered
as it was impractical to implement the EIS algorithm in the Stan language.
For each of the three models, the LD algorithm is simulated for 1,500
iterations, where the draws from the first 500 burn-in iterations
are discarded. Stan uses (the default) 2,000 iterations with 1,000
burn-in steps also used for automatic tuning of the integrator step 
size and the mass matrix. The reported computing
times are for the 1,000 sampling iterations for both methods. Further
details for the different example models, including prior assumptions and details related to the
Newton iterations for the Laplace maps, 
are found in the supplementary material, Section~\ref{sec:Details-related-to-sim-study}.

\subsection{Stochastic Volatility Model}

The first example model is the discrete-time stochastic volatility
(SV) model for financial returns given by \citep{Taylor1986} 
\begin{align}
y_{t} & =\exp(x_{t}/2)e_{t},\quad e_{t}\sim\mbox{iid}\ N(0,1),\;t=1,\dots,D,\label{eq:SV_1}\\
x_{t} & =\gamma+\delta x_{t-1}+\nu\eta_{t},\quad\eta_{t}\sim\mbox{iid}\ N(0,1),\;t=2,\dots,D,\label{eq:SV_2}
\end{align}
where $y_{t}$ is the return observed on day $t$, $x_{t}$ is the
latent log-volatility with initial condition $x_{1}\sim N(\gamma/[1-\delta],\nu^{2}/[1-\delta^{2}])$,
while $e_{t}$ and $\eta_{t}$ are mutually independent innovations.
The data consists of daily log-returns on
the U.S. dollar against the U.K. Pound Sterling from October 1, 1981
to June 28, 1985 with $D=945$. 

Under this SV model the data density $p(y_{t}|x_{t})={\cal {N}}(y_{t}|0,\exp\{x_{t}\})$
is fairly uninformative about the states $x_{t}$, with a Fisher information
(w.r.t. $x_{t}$) which is independent of $\boldsymbol{\theta}$ and
given by $-E[\nabla_{x_{t}}^{2}\log p(y_{t}|x_{t})]=1/2$, whereas
the states are fairly volatile under typical estimates for $\boldsymbol{\theta}$.
This low signal-to-noise ratio together with a shape of the data density
which is independent of the parameters implies that the conditional
posterior of the innovations $\boldsymbol{\eta}$ given $\boldsymbol{\theta}$
are close to a normal distribution regardless of $\boldsymbol{\theta}$,
leading to a correspondingly well-behaved joint posterior of $\boldsymbol{\theta}$
and $\boldsymbol{\eta}$. Hence, this represents a scenario where
the Stan-Prior sampling on the joint space of $\boldsymbol{\theta}$
and $\boldsymbol{\eta}$ used as a benchmark can be expected to exhibit
a comparably good performance.

For the Fisher transport map method, $\mathbf{G}_{F}=\mathbf{G_{\boldsymbol{\theta},x}}+\mathbf{G_{\boldsymbol{\theta},y|\mathbf{x}}}$,
and as suggested by Table 4 of \citet{1806.02068}, we set $\mathbf{h}_{F}=\mathbf{0}_{d}$.

\begin{table}
\centering{}{\scriptsize{}}%
\begin{tabular}{llllcllcllclllll}
\hline 
 &  & \multicolumn{2}{c}{{\scriptsize{}LD-EIS}} &  & \multicolumn{2}{c}{{\scriptsize{}Stan-Prior}} &  & \multicolumn{2}{c}{{\scriptsize{}LD-Laplace}} &  & \multicolumn{2}{c}{{\scriptsize{}Stan-Laplace}} &  & \multicolumn{2}{l}{{\scriptsize{}Stan-Fisher}}\tabularnewline
\cline{3-4} \cline{4-4} \cline{6-7} \cline{7-7} \cline{9-10} \cline{10-10} \cline{12-16} \cline{13-16} \cline{14-16} \cline{15-16} \cline{16-16} 
 &  & {\scriptsize{}Min } & {\scriptsize{}Mean } &  & {\scriptsize{}Min } & {\scriptsize{}Mean } &  & {\scriptsize{}Min } & {\scriptsize{}Mean } &  & {\scriptsize{}Min } & {\scriptsize{}Mean } &  & {\scriptsize{}Min } & {\scriptsize{}Mean }\tabularnewline
\hline 
 & {\scriptsize{}CPU time (s) } & {\scriptsize{}276.5 } & {\scriptsize{}278 } &  & {\scriptsize{}12.4 } & {\scriptsize{}15 } &  & {\scriptsize{}10.6 } & {\scriptsize{}10.6 } &  & {\scriptsize{}9.7 } & {\scriptsize{}16.7 } &  & {\scriptsize{}6.1 } & {\scriptsize{}7.6 }\tabularnewline
{\scriptsize{}$\gamma$ } &  &  &  &  &  &  &  &  &  &  &  &  &  &  & \tabularnewline
 & {\scriptsize{}Post. mean } &  & {\scriptsize{}-0.021 } &  &  & {\scriptsize{}-0.021 } &  &  & {\scriptsize{}-0.021 } &  &  & {\scriptsize{}-0.021 } &  &  & {\scriptsize{}-0.021 }\tabularnewline
 & {\scriptsize{}Post. std. } &  & {\scriptsize{}0.012 } &  &  & {\scriptsize{}0.01 } &  &  & {\scriptsize{}0.011 } &  &  & {\scriptsize{}0.011 } &  &  & {\scriptsize{}0.011 }\tabularnewline
 & {\scriptsize{}ESS } & {\scriptsize{}201 } & {\scriptsize{}337 } &  & {\scriptsize{}237 } & {\scriptsize{}348 } &  & {\scriptsize{}275 } & {\scriptsize{}354 } &  & {\scriptsize{}268 } & {\scriptsize{}494 } &  & {\scriptsize{}218 } & {\scriptsize{}321 }\tabularnewline
 & {\scriptsize{}ESS/s } & {\scriptsize{}0.7 } & {\scriptsize{}1.2 } &  & {\scriptsize{}18.1 } & {\scriptsize{}23.5 } &  & {\scriptsize{}25.7 } & {\scriptsize{}33.3 } &  & {\scriptsize{}16.7 } & {\scriptsize{}37.2 } &  & {\scriptsize{}5.6 } & {\scriptsize{}27 }\tabularnewline
{\scriptsize{}$\delta$ } &  &  &  &  &  &  &  &  &  &  &  &  &  &  & \tabularnewline
 & {\scriptsize{}Post. mean } &  & {\scriptsize{}0.98 } &  &  & {\scriptsize{}0.98 } &  &  & {\scriptsize{}0.98 } &  &  & {\scriptsize{}0.98 } &  &  & {\scriptsize{}0.98 }\tabularnewline
 & {\scriptsize{}Post. std. } &  & {\scriptsize{}0.01 } &  &  & {\scriptsize{}0.01 } &  &  & {\scriptsize{}0.01 } &  &  & {\scriptsize{}0.01 } &  &  & {\scriptsize{}0.01 }\tabularnewline
 & {\scriptsize{}ESS } & {\scriptsize{}269 } & {\scriptsize{}380 } &  & {\scriptsize{}192 } & {\scriptsize{}309 } &  & {\scriptsize{}320 } & {\scriptsize{}363 } &  & {\scriptsize{}290 } & {\scriptsize{}423 } &  & {\scriptsize{}239 } & {\scriptsize{}319 }\tabularnewline
 & {\scriptsize{}ESS/s } & {\scriptsize{}1 } & {\scriptsize{}1.4 } &  & {\scriptsize{}15.3 } & {\scriptsize{}20.6 } &  & {\scriptsize{}30.1 } & {\scriptsize{}34.1 } &  & {\scriptsize{}13.9 } & {\scriptsize{}32 } &  & {\scriptsize{}5 } & {\scriptsize{}27.2 }\tabularnewline
{\scriptsize{}$v$ } &  &  &  &  &  &  &  &  &  &  &  &  &  &  & \tabularnewline
 & {\scriptsize{}Post. mean } &  & {\scriptsize{}0.15 } &  &  & {\scriptsize{}0.15 } &  &  & {\scriptsize{}0.15 } &  &  & {\scriptsize{}0.15 } &  &  & {\scriptsize{}0.15 }\tabularnewline
 & {\scriptsize{}Post. std. } &  & {\scriptsize{}0.03 } &  &  & {\scriptsize{}0.03 } &  &  & {\scriptsize{}0.03 } &  &  & {\scriptsize{}0.03 } &  &  & {\scriptsize{}0.03 }\tabularnewline
 & {\scriptsize{}ESS } & {\scriptsize{}363 } & {\scriptsize{}503 } &  & {\scriptsize{}243 } & {\scriptsize{}332 } &  & {\scriptsize{}360 } & {\scriptsize{}512 } &  & {\scriptsize{}274 } & {\scriptsize{}431 } &  & {\scriptsize{}226 } & {\scriptsize{}293 }\tabularnewline
 & {\scriptsize{}ESS/s } & {\scriptsize{}1.3 } & {\scriptsize{}1.8 } &  & {\scriptsize{}16.7 } & {\scriptsize{}23 } &  & {\scriptsize{}33.9 } & {\scriptsize{}48.1 } &  & {\scriptsize{}14.1 } & {\scriptsize{}32.8 } &  & {\scriptsize{}3.8 } & {\scriptsize{}25.6 }\tabularnewline
\hline 
\end{tabular}{\scriptsize{} \caption{\label{tab:SV}Simulation study results for the SV model~(\ref{eq:SV_1},\ref{eq:SV_2}).
ESS corresponds to the effective sample size (out of 1,000 iterations)
and ESS/s is the number of effective samples produced per second of
computing time. The columns ``Min'', ``Mean'' correspond to the
minimum, mean across 8 independent replicas of the experiment. Burn-in
iterations are not included in the reported CPU times. The tuning
parameters are: LD-EIS: $J=2$, $r=6$, $\varepsilon=0.4$ and $L=4$.
LD-Laplace: $K=2$, $\varepsilon=0.4$ and $L=4$. Stan-Laplace: $K=0$.}
}
\end{table}

Table~\ref{tab:SV} shows the HMC posterior mean and standard deviation
for the parameters, which are sample averages computed from 8 independent
replications. It also reports the effective sample size (ESS) \citep{Geyer1992}
and the ESS per second of CPU time (ESS/s), where the latter will
be the main performance measure (provided of course that the MCMC
method properly explores the target distribution) considered here.
Several settings of the tuning parameters (i.e. some subset of $r$,
$J$, $K$, and $L$) where considered, and the presented results
are the best considered, in terms of ESS/s. Table~\ref{tab:SV} indicates
firstly that all five methods produce a good exploration of the target
distribution with posterior moments being essentially the same. For
the Stan-based methods, there is substantial variation in the CPU
times due to variation in the automatic tuning of the integrator step
size $\varepsilon$ over the replica. Judging from the ESS values, on
average there is not much to be gained from introducing the Laplace
approximation- and EIS-based transport map for this model. This finding mirrors
to some extent what was found by \citet[Section 5.2]{1806.02068},
and is also as expected since the observations carry very little information
regarding the states. In terms of ESS/s, there is no uniform winner,
but the computational overhead of locating the EIS importance density
is clearly not worthwhile for this model, relative to the computationally
cheaper Laplace- and Fisher transport maps. 

\subsection{Gamma Model for Realized Volatilities\label{subsec:Gamma-Model-for}}

The second example model is a dynamic state-space model for the realized
variance of asset returns \citep[see, e.g.,][and references therein]{Golosnoy2012}.
It has the form 
\begin{align}
y_{t} & =\beta\exp(x_{t})e_{t},\quad e_{t}\sim\mbox{iid}\ G(1/\tau,\tau),\;t=1,\dots,D,\label{eq:Gamma_1}\\
x_{t} & =\delta x_{t-1}+\nu\eta_{t},\quad\eta_{t}\sim\mbox{iid}\ N(0,1),\;t=2,\dots,D,\label{eq:Gamma_2}
\end{align}
where $y_{t}$ is the daily realized variance measuring the latent
integrated variance $\beta\exp(x_{t})$, and $G(1/\tau,\tau)$ denotes
a Gamma-distribution for $e_{t}$ normalized such that $E(e_{t})=1$
and Var$(e_{t})=\tau$. The innovations $e_{t}$ and $\eta_{t}$ are
independent and the initial condition for the log-variance is $x_{1}\sim N(0,\nu^{2}/[1-\delta^{2}])$.
This Gamma volatility model is applied to a data set consisting of
$D=2,514$ observations of the daily realized variance for the American
Express stock (more information concerning the data is given in Section~
\ref{sec:Realistic-application}; $y_{t}$ here is identical to the
1,1-element of realized covariance matrices $\mathbf{Y}_{t}$).

In contrast to the SV model, this Gamma model applied to the realized
variance data has both a considerably higher signal-to-noise ratio
and a shape of the data density $x_{t}\mapsto p(y_{t}|x_{t},\boldsymbol{\theta})$
which depends on the parameters. In particular, the Fisher information
of its data density with respect to $x_{t}$ is $1/\tau$ with an
estimate of $\tau\simeq0.13$ (see Table~\ref{tab:Gamma}), while
the estimated volatility of the states is roughly as large as under
the SV model. Hence, it can be expected that the conditional posterior
of the innovations $\boldsymbol{\eta}$ given $\boldsymbol{\theta}$
deviates distinctly from a Gaussian form and exhibits nonlinear dependence
on $\boldsymbol{\theta}$, which makes the Gamma model a more challenging
scenario for the Stan-Prior benchmark than the SV model.

\begin{table}[t]
\centering{}%
\begin{tabular}{llllcllcllcll}
\hline 
 &  & \multicolumn{2}{c}{{\scriptsize{}LD-EIS}} &  & \multicolumn{2}{c}{{\scriptsize{}Stan-Prior}} &  & \multicolumn{2}{c}{{\scriptsize{}LD-Laplace}} &  & \multicolumn{2}{c}{{\scriptsize{}Stan-Laplace}} \tabularnewline
\cline{3-4} \cline{4-4} \cline{6-7} \cline{7-7} \cline{9-10} \cline{10-10} \cline{12-13} 
 &  & {\scriptsize{}Min } & {\scriptsize{}Mean } &  & {\scriptsize{}Min } & {\scriptsize{}Mean } &  & {\scriptsize{}Min } & {\scriptsize{}Mean } &  & {\scriptsize{}Min } & {\scriptsize{}Mean } \tabularnewline
\hline 
 & {\scriptsize{}CPU time (s) } & {\scriptsize{}935.4 } & {\scriptsize{}938.1 } &  & {\scriptsize{}150.5 } & {\scriptsize{}171.1 } &  & {\scriptsize{}50.9 } & {\scriptsize{}51.1 } &  & {\scriptsize{}40.8 } & {\scriptsize{}62 } \tabularnewline
{\scriptsize{}$\tau$ } \tabularnewline
 & {\scriptsize{}Post. mean } &  & {\scriptsize{}0.13 } &  &  & {\scriptsize{}0.13 } &  &  & {\scriptsize{}0.13 } &  &  & {\scriptsize{}0.13 }   \tabularnewline
 & {\scriptsize{}Post. std. } &  & {\scriptsize{}0.006 } &  &  & {\scriptsize{}0.006 } &  &  & {\scriptsize{}0.006 } &  &  & {\scriptsize{}0.006 } \tabularnewline
 & {\scriptsize{}ESS } & {\scriptsize{}1000 } & {\scriptsize{}1000 } &  & {\scriptsize{}194 } & {\scriptsize{}238 } &  & {\scriptsize{}1000 } & {\scriptsize{}1000 } &  & {\scriptsize{}623 } & {\scriptsize{}873 } \tabularnewline
 & {\scriptsize{}ESS/s } & {\scriptsize{}1.1 } & {\scriptsize{}1.1 } &  & {\scriptsize{}1.1 } & {\scriptsize{}1.4 } &  & {\scriptsize{}19.5 } & {\scriptsize{}19.6 } &  & {\scriptsize{}10.1 } & {\scriptsize{}15.2 } \tabularnewline
{\scriptsize{}$\beta$ } \tabularnewline
 & {\scriptsize{}Post. mean } &  & {\scriptsize{}2.7 } &  &  & {\scriptsize{}2.8 } &  &  & {\scriptsize{}2.5 } &  &  & {\scriptsize{}2.8 } \tabularnewline
 & {\scriptsize{}Post. std. } &  & {\scriptsize{}0.8 } &  &  & {\scriptsize{}1 } &  &  & {\scriptsize{}0.8 } &  &  & {\scriptsize{}0.9 } \tabularnewline
 & {\scriptsize{}ESS } & {\scriptsize{}460 } & {\scriptsize{}542 } &  & {\scriptsize{}65 } & {\scriptsize{}281 } &  & {\scriptsize{}216 } & {\scriptsize{}568 } &  & {\scriptsize{}103 } & {\scriptsize{}505 } \tabularnewline
 & {\scriptsize{}ESS/s } & {\scriptsize{}0.5 } & {\scriptsize{}0.6 } &  & {\scriptsize{}0.4 } & {\scriptsize{}1.7 } &  & {\scriptsize{}4.2 } & {\scriptsize{}11.1 } &  & {\scriptsize{}2.5 } & {\scriptsize{}8.3 } \tabularnewline
{\scriptsize{}$\delta$ }  \tabularnewline
 & {\scriptsize{}Post. mean } &  & {\scriptsize{}0.98 } &  &  & {\scriptsize{}0.98 } &  &  & {\scriptsize{}0.98 } &  &  & {\scriptsize{}0.98 } \tabularnewline
 & {\scriptsize{}Post. std. } &  & {\scriptsize{}0.004 } &  &  & {\scriptsize{}0.004 } &  &  & {\scriptsize{}0.004 } &  &  & {\scriptsize{}0.004 } \tabularnewline
 & {\scriptsize{}ESS } & {\scriptsize{}497 } & {\scriptsize{}641 } &  & {\scriptsize{}207 } & {\scriptsize{}282 } &  & {\scriptsize{}384 } & {\scriptsize{}685 } &  & {\scriptsize{}382 } & {\scriptsize{}719 } \tabularnewline
 & {\scriptsize{}ESS/s } & {\scriptsize{}0.5 } & {\scriptsize{}0.7 } &  & {\scriptsize{}1.3 } & {\scriptsize{}1.7 } &  & {\scriptsize{}7.5 } & {\scriptsize{}13.4 } &  & {\scriptsize{}8.7 } & {\scriptsize{}11.9 } \tabularnewline
{\scriptsize{}$\nu$ } \tabularnewline
 & {\scriptsize{}Post. mean } &  & {\scriptsize{}0.22 } &  &  & {\scriptsize{}0.22 } &  &  & {\scriptsize{}0.22 } &  &  & {\scriptsize{}0.22 } \tabularnewline
 & {\scriptsize{}Post. std. } &  & {\scriptsize{}0.01 } &  &  & {\scriptsize{}0.01 } &  &  & {\scriptsize{}0.01 } &  &  & {\scriptsize{}0.01 }\tabularnewline
 & {\scriptsize{}ESS } & {\scriptsize{}827 } & {\scriptsize{}976 } &  & {\scriptsize{}139 } & {\scriptsize{}178 } &  & {\scriptsize{}1000 } & {\scriptsize{}1000 } &  & {\scriptsize{}416 } & {\scriptsize{}785 } \tabularnewline
 & {\scriptsize{}ESS/s } & {\scriptsize{}0.9 } & {\scriptsize{}1 } &  & {\scriptsize{}0.6 } & {\scriptsize{}1.1 } &  & {\scriptsize{}19.5 } & {\scriptsize{}19.6 } &  & {\scriptsize{}8.3 } & {\scriptsize{}13.4 } \tabularnewline
\hline 
\end{tabular} \caption{\label{tab:Gamma}Simulation study results for the Gamma model~(\ref{eq:Gamma_1},\ref{eq:Gamma_2}).
ESS corresponds to the effective sample size (out of 1,000 iterations)
and ESS/s is the number of effective samples produced per second of
computing time. The columns ``Min'', ``Mean'' correspond to the
minimum, mean across 8 independent replicas of the experiment. Burn-in
iterations are not included in the reported CPU times. The tuning
parameters are: LD-EIS: $J=2$, $r=5$, $\varepsilon=0.64$ and $L=3$,
LD-Laplace: $K=1$, $\varepsilon=0.64$ and $L=3$. Stan-Laplace:
$K=0$. Notice that Stan-Fisher and Stan-Laplace coincide in this case.}
\end{table}

The same initial guess $\mathbf{h}^{(0)}$ in the Laplace scaling
as for the SV model above was applied, and also here $\mathbf{G}_{F}$
coincides with $\mathbf{G_{\boldsymbol{\theta},x}}+\mathbf{G_{\boldsymbol{\theta},y|\mathbf{x}}}$.
Choosing $\mathbf{h}_{F}=\mathbf{0}$ leads to poor results, and we 
therefore set $\mathbf{h}_{F}$ equal to~(\ref{eq:laplace_guess}) \citep[see also][Equation 20]{1806.02068}.  
Consequently, Stan-Fisher coincides with Stan-Laplace, $K=0$ (which was also found to be the
optimal Stan-Laplace method in this situation).
The remaining experiment setup is also identical to that for the SV
model, and the results are given in Table~\ref{tab:Gamma}. Stan-Prior
produces substantially lower ESSes than the EIS- and Laplace methods,
which we attribute to the failure to take the higher information content
from the observations into account in the transport map. LD-Laplace and
Stan-Laplace are the winners in terms of ESS/s and again it is not
beneficial to opt for the presumably more accurate and expensive EIS-transport map
over the cruder and computationally faster Laplace-approximation.

\subsection{Constant Elasticity of Variance Diffusion Model}

The last example model is a time-discretized version of the constant
elasticity of variance (CEV) diffusion model for short-term interest
rates \citep{Chan1992}, extended by a measurement error to account
for microstructure noise \citep{Ait-Sahalia1999,Kleppe2016}. The
resulting model for the interest rate $y_{t}$ observed at day $t$
with a corresponding latent state $x_{t}>0$ , is described as 
\begin{align}
y_{t} & =x_{t}+\sigma_{y}e_{t},\quad e_{t}\sim\mbox{iid}\ N(0,1),\;t=1,\dots,D,\label{eq:CEV_1}\\
x_{t} & =x_{t-1}+\Delta(\alpha-\beta x_{t-1})+\sigma_{x}x_{t-1}^{\gamma}\sqrt{\Delta}\eta_{t},\quad\eta_{t}\sim\mbox{iid}\ N(0,1),\;t=2,\dots,D,\label{eq:CEV_2}
\end{align}
where $e_{t}$ and $\eta_{t}$ are mutually independent and $\Delta=1/252$.
The parameters are $\boldsymbol{\theta}=(\alpha,\beta,\gamma,\sigma_{x},\sigma_{y})$
and the initial condition $x_{1}\sim N(y_{1},0.01^{2})$. The data
consist of $D=3,082$ daily 7-day Eurodollar deposit spot rates from
January 2, 1983 to February 25, 1995 (see \citealp{Ait-Sahalia1996}
for a description of this data set). 

The estimated standard
deviation of the noise component $\sigma_{y}$ is very small with
an estimate of 0.0005 (see Table~\ref{tab:CEV}) so that the data
density $x_{t}\mapsto p(y_{t}|x_{t},\boldsymbol{\theta})$ is strongly
peaked at $x_{t}=y_{t}$ and by far more informative about $x_{t}$
than in the SV- and Gamma model with a Fisher information given by
$1/\sigma_{y}^{2}$. Also, the volatility of the states is not constant
and depends, unlike in the previous models, nonlinearly on the level
of the states. As a result, the posterior of $\boldsymbol{\eta}$ and
$\boldsymbol{\theta}$ strongly deviates from being Gaussian. Consequently,
Stan-Prior fails to produce meaningful results and is therefore not
reported on. Moreover, since the prior on $\mathbf{x}$ is nonlinear and its precision matrix
does not seem to have closed-form,
Fisher-scaling is not feasible.

\begin{table}[t]
\centering{}{\footnotesize{}}%
\begin{tabular}{llllcllcll}
\hline 
 &  & \multicolumn{2}{c}{{\scriptsize{}LD-EIS}} &  & \multicolumn{2}{c}{{\scriptsize{}LD-Laplace}} &  & \multicolumn{2}{c}{{\scriptsize{}Stan-Laplace}}\tabularnewline
\cline{3-4} \cline{4-4} \cline{6-7} \cline{7-7} \cline{9-10} \cline{10-10} 
 &  & {\scriptsize{}Min } & {\scriptsize{}Mean } &  & {\scriptsize{}Min } & {\scriptsize{}Mean } &  & {\scriptsize{}Min } & {\scriptsize{}Mean }\tabularnewline
\hline 
 & {\scriptsize{}CPU time (s) } & {\scriptsize{}615.6 } & {\scriptsize{}618.8 } &  & {\scriptsize{}60.3 } & {\scriptsize{}60.6 } &  & {\scriptsize{}482.2 } & {\scriptsize{}515.7 }\tabularnewline
{\scriptsize{}$\alpha$ } &  &  &  &  &  &  &  &  & \tabularnewline
 & {\scriptsize{}Post. mean } &  & {\scriptsize{}0.01 } &  &  & {\scriptsize{}0.01 } &  &  & {\scriptsize{}0.01 }\tabularnewline
 & {\scriptsize{}Post. std. } &  & {\scriptsize{}0.01 } &  &  & {\scriptsize{}0.01 } &  &  & {\scriptsize{}0.01 }\tabularnewline
 & {\scriptsize{}ESS } & {\scriptsize{}869 } & {\scriptsize{}984 } &  & {\scriptsize{}876 } & {\scriptsize{}972 } &  & {\scriptsize{}1000 } & {\scriptsize{}1000 }\tabularnewline
 & {\scriptsize{}ESS/s } & {\scriptsize{}1.4 } & {\scriptsize{}1.6 } &  & {\scriptsize{}14.5 } & {\scriptsize{}16 } &  & {\scriptsize{}1.9 } & {\scriptsize{}1.9 }\tabularnewline
{\scriptsize{}$\beta$ } &  &  &  &  &  &  &  &  & \tabularnewline
 & {\scriptsize{}Post. mean } &  & {\scriptsize{}0.17 } &  &  & {\scriptsize{}0.17 } &  &  & {\scriptsize{}0.17 }\tabularnewline
 & {\scriptsize{}Post. std. } &  & {\scriptsize{}0.17 } &  &  & {\scriptsize{}0.17 } &  &  & {\scriptsize{}0.17 }\tabularnewline
 & {\scriptsize{}ESS } & {\scriptsize{}707 } & {\scriptsize{}963 } &  & {\scriptsize{}745 } & {\scriptsize{}957 } &  & {\scriptsize{}1000 } & {\scriptsize{}1000 }\tabularnewline
 & {\scriptsize{}ESS/s } & {\scriptsize{}1.1 } & {\scriptsize{}1.6 } &  & {\scriptsize{}12.4 } & {\scriptsize{}15.8 } &  & {\scriptsize{}1.9 } & {\scriptsize{}1.9 }\tabularnewline
{\scriptsize{}$\gamma$ } &  &  &  &  &  &  &  &  & \tabularnewline
 & {\scriptsize{}Post. mean } &  & {\scriptsize{}1.18 } &  &  & {\scriptsize{}1.18 } &  &  & {\scriptsize{}1.18 }\tabularnewline
 & {\scriptsize{}Post. std. } &  & {\scriptsize{}0.06 } &  &  & {\scriptsize{}0.06 } &  &  & {\scriptsize{}0.06 }\tabularnewline
 & {\scriptsize{}ESS } & {\scriptsize{}759 } & {\scriptsize{}957 } &  & {\scriptsize{}1000 } & {\scriptsize{}1000 } &  & {\scriptsize{}631 } & {\scriptsize{}852 }\tabularnewline
 & {\scriptsize{}ESS/s } & {\scriptsize{}1.2 } & {\scriptsize{}1.5 } &  & {\scriptsize{}16.4 } & {\scriptsize{}16.5 } &  & {\scriptsize{}1.3 } & {\scriptsize{}1.6 }\tabularnewline
{\scriptsize{}$\sigma_{x}$ } &  &  &  &  &  &  &  &  & \tabularnewline
 & {\scriptsize{}Post. mean } &  & {\scriptsize{}0.41 } &  &  & {\scriptsize{}0.41 } &  &  & {\scriptsize{}0.41 }\tabularnewline
 & {\scriptsize{}Post. std. } &  & {\scriptsize{}0.06 } &  &  & {\scriptsize{}0.06 } &  &  & {\scriptsize{}0.06 }\tabularnewline
 & {\scriptsize{}ESS } & {\scriptsize{}769 } & {\scriptsize{}946 } &  & {\scriptsize{}1000 } & {\scriptsize{}1000 } &  & {\scriptsize{}650 } & {\scriptsize{}890 }\tabularnewline
 & {\scriptsize{}ESS/s } & {\scriptsize{}1.2 } & {\scriptsize{}1.5 } &  & {\scriptsize{}16.4 } & {\scriptsize{}16.5 } &  & {\scriptsize{}1.3 } & {\scriptsize{}1.7 }\tabularnewline
{\scriptsize{}$\sigma_{y}$ } &  &  &  &  &  &  &  &  & \tabularnewline
 & {\scriptsize{}Post. mean } &  & {\scriptsize{}0.0005 } &  &  & {\scriptsize{}0.0005 } &  &  & {\scriptsize{}0.0005 }\tabularnewline
 & {\scriptsize{}Post. std. } &  & {\scriptsize{}0.00002 } &  &  & {\scriptsize{}0.00002 } &  &  & {\scriptsize{}0.00002 }\tabularnewline
 & {\scriptsize{}ESS } & {\scriptsize{}769 } & {\scriptsize{}963 } &  & {\scriptsize{}1000 } & {\scriptsize{}1000 } &  & {\scriptsize{}1000 } & {\scriptsize{}1000 }\tabularnewline
 & {\scriptsize{}ESS/s } & {\scriptsize{}1.2 } & {\scriptsize{}1.6 } &  & {\scriptsize{}16.4 } & {\scriptsize{}16.5 } &  & {\scriptsize{}1.9 } & {\scriptsize{}1.9 }\tabularnewline
\hline 
\end{tabular} \caption{\label{tab:CEV}Simulation study results for the CEV model~(\ref{eq:CEV_1},\ref{eq:CEV_2}).
ESS corresponds to the effective sample size (out of 1,000 iterations)
and ESS/s is the number of effective samples produced per second of
computing time. The columns ``Min'', ``Mean'' correspond to the
minimum, mean across 8 independent replicas of the experiment. Burn-in
iterations are not included in the reported CPU times. The tuning
parameters are: LD-EIS: $J=1$, $r=7$, $\epsilon=0.57$ and $L=3$.
LD-Laplace: $K=2$, $\epsilon=0.57$ and $L=3$, Stan-Laplace: $K=1$.}
\end{table}

Table~\ref{tab:CEV} reports results for LD-EIS, LD-Laplace and Stan-Laplace,
and it is seen that all three methods produce reliable results. In
terms of ESS per computing time, the LD-Laplace is a factor 5-10 faster
than the other methods, where the difference between LD-Laplace and
Stan-Laplace is due to the substantially higher number of integrator
steps required for Stan-Laplace. 

The same model and data set was also considered by \citet[Section 5]{Kleppe2018},
who compare the modified Cholesky Riemann manifold HMC algorithm and
a Gibbs sampling procedure. Both methods were implemented
in C++ and thus the orders of magnitude of produced ESS per computing
time are comparable to the present situation. It is seen
that for the ``most difficult'' parameters $\gamma,\;\sigma_{x}$,
the proposed methodology is roughly two order of magnitude faster
than the Riemann manifold HMC method and roughly three orders of magnitude
faster than the Gibbs sampler.

\subsection{Summary from simulation experiment}

For models with higher signal-to-noise ratios than the SV model, the
proposed methodology produces large speedups (or makes challenging
models feasible as for the CEV model) relative to the benchmarks,
even if the per evaluation cost of the modified target is higher than
in the default parameterization. For the considered models, the EIS
transport map is not competitive relative to the Laplace approximation
counterpart due to the relatively higher computational cost. For the
Laplace-based methods, it is seen that relatively few Newton iterations
is optimal in an ESS per computing time perspective. Overall, and
very much in line with \citet{1806.02068}, this is indicative that
rather crude representations of the location and scale of $p(\mathbf{x}|\mathbf{y},\boldsymbol{\theta})$
are sufficient. Moreover, this latter observation ties in with the
second point discussed in Section~\ref{subsec:Relation-to-importance}:
Due to the thin-tailed Gaussian distribution entering explicitly
in representation~(\ref{eq:t-target-2}) of the modified target, the
importance sampling rule of thumb that you should seek high-fidelity
approximations to $p(\mathbf{x}|\mathbf{y},\boldsymbol{\theta})$
as the importance density is less relevant in the present situation. 

With respect to the choice of integrator, it is seen that the LD-integrator
and the leapfrog-integrator-based Stan produces similar raw ESSes,
but that that the LD-integrator in general requires non-trivially
fewer integration steps to accomplish this. E.g., the reported (automatically
tuned) Stan-Laplace results for the CEV model required on average
63 leapfrog steps whereas the corresponding (manually tuned) number
for LD-Laplace was 3. For the two other models, the performance of
the LD integrator is roughly on par with Stan when Laplace scaling
was employed. Further, the LD integrator generally needs more refined
Laplace maps (higher $K$) to work satisfactory, whereas under Stan,
more crude Laplace transport maps are permissible. 

\section{High-dimensional application\label{sec:Realistic-application}}

\subsection{Model\label{subsec:Wishart-Model}}

To illustrate the proposed methodology in a high-dimensional situation, we consider
the dynamic inverted Wishart model for realized covariance matrices
proposed in \citet[Section 6]{grothe2017}. More specifically, for
a time series of $r\times r$ symmetric positive definite observed
realized covariance matrices $\mathbf{Y}_{t},\;t=1,\dots,D$, the
observations are modeled conditionally inverse-Wishart distributed,
\begin{equation}
p(\mathbf{Y}_{t}|\boldsymbol{\Sigma}_{t},\nu)\propto|\mathbf{Y}_{t}|^{-\frac{\nu+r+1}{2}}\exp\left(-\frac{1}{2}\text{tr}\left(\boldsymbol{\Sigma}_{t}\mathbf{Y}_{t}^{-1}\right)\right),\label{eq:wishart_mod_1}
\end{equation}
so that $E(\mathbf{Y}_{t})=(\nu+r+1)^{-1}\boldsymbol{\Sigma}_{t}$.
Here, the degrees of freedom $\nu>r+1$ is a parameter, and $\boldsymbol{\Sigma}_{t}$
is a (latent) time-varying scale matrix, given by
\[
\boldsymbol{\Sigma}_{t}=\mathbf{H}\mathbf{D}_{t}\mathbf{H}^{T},\;\mathbf{D}_{t}=\text{diag}(\exp(x_{1,t}),\dots,\exp(x_{r,t})),
\]
where $\boldsymbol{H}$ is a lower triangular matrix with ones along
the main diagonal and unrestricted parameters $h_{i,j},i>j,1\leq j<r$
below the main diagonal. Moreover, $\mathbf{x}_{s}=\{x_{s,t}\}_{t=1}^{D},\;s=1,\dots,r$
are latent Gaussian AR(1) processes
\begin{align}
x_{s,t} & =\mu_{s}+\delta_{s}(x_{s,t-1}-\mu_{s})+\sigma_{s}\eta_{s,t},\;t=2,\dots,D,\;s=1,\dots,r,\label{eq:wishart_mod_2}\\
x_{s,1} & =\mu_{s}+\frac{\sigma_{s}}{\sqrt{1-\delta_{s}^{2}}}\eta_{s,1},\;s=1,\dots,r\label{eq:wishart_mod_3}
\end{align}
where $\eta_{s,t}\sim\text{iid }N(0,1),\;t=1,\dots,D,\;s=1,\dots,r$.
In total, the model contains $1+3r+r(r-1)/2$ parameters $\boldsymbol{\theta}=(\nu,\mu_{1:r},\delta_{1:r},\sigma_{1:r},h_{2:r,1},h_{3:r,2},\dots,h_{r,r-1})$.
Further details concerning the model specification and priors can
be found in the supplementary material (Section~\ref{sec:Details-related-to}).

A fortunate property of this model is that the conditional posterior
of the latent states are independent over $s$, i.e. $p(\mathbf{x}_{1:r}|\boldsymbol{\theta},Y_{1:D})=\prod_{s=1}^{r}p(\mathbf{x}_{s}|\boldsymbol{\theta},Y_{1:D})$.
This implies that the transport map for $\mathbf{x}$ also may be
split into $r$ individual transport maps, say $\mathbf{x}_{s}=\gamma_{\boldsymbol{\theta},s}(\mathbf{u}_{s}),\;\mathbf{u}_{s}=\{u_{s,t}\}_{t=1}^{D},\;s=1,\dots,r,$
without losing fidelity. The (combined) transport map becomes $\gamma_{\boldsymbol{\theta}}(\mathbf{u})=[\left(\gamma_{\boldsymbol{\theta},1}(\mathbf{u}_{1})\right)^{T},\dots,\left(\gamma_{\boldsymbol{\theta},r}(\mathbf{u}_{r})\right)^{T}]^{T}$,
where $\mathbf{u}=[\mathbf{u}_{1}^{T},\dots,\mathbf{u}_{r}^{T}]^{T}$,
and in particular $|\nabla_{\mathbf{u}}\gamma_{\boldsymbol{\theta}}(\mathbf{u})|=\prod_{s=1}^{r}|\nabla_{\mathbf{u}_{s}}\gamma_{\boldsymbol{\theta},s}(\mathbf{u}_{s})|$
due to the block-diagonal nature of the Jacobian of $\gamma_{\boldsymbol{\theta}}$. 

Further, each of the factors of the conditional posterior have a shape
corresponding that of a state-space model with univariate state-process $\mathbf x_s$:
\begin{equation}
p(\mathbf{x}_{s}|\boldsymbol{\theta},Y_{1:D})\propto
 p(\mathbf{x}_{s}|\boldsymbol{\theta})\prod_{t=1}^{D}\exp\left(\frac{\nu}{2}x_{s,t}-\frac{\tilde{y}_{s,t}}{2}\exp(x_{s,t})\right),
 \;\tilde{y}_{s,t}=\left(\mathbf{H}_{1:s,s}\right)^{T}\mathbf{Y}_{t}^{-1}\mathbf{H}_{1:s,s},\;s=1,\dots,r.\label{eq:wishart-marg-target}
\end{equation}
Thus, individual transport maps $\gamma_{\boldsymbol{\theta},s}$
may be constructed to target~(\ref{eq:wishart-marg-target}) as described
in the previous Sections. In particular, individual Laplace approximation-based
maps, $\gamma_{\boldsymbol{\theta},s}$, involve only tri-diagonal
Cholesky factorizations. It is, however, worth noticing that the proposed
methodology does not rely on such a conditional independence structure
in order to be applicable per se. 

The observed Fisher information (w.r.t. $x_{s,t}$) of the marginal
``measurement densities'' $\propto\exp(\frac{\nu}{2}x_{s,t}-\frac{\tilde{y}_{s,t}}{2}\exp(x_{s,t}))$
equals $\nu/2$, with an estimate of $\nu\simeq33.6$ for the data
set considered here (see Table~\ref{tab:Posterior-mean-and-wishart}
in supplementary material). Thus, the signal to noise ratio here is
similar to that of the Gamma model considered in section~\ref{subsec:Gamma-Model-for}.
As the LD- and Stan- results are similar for the Gamma model,
we consider only Stan for this model, as it entails only a few dozen
lines of Stan code and tuning is fully automated. 
EIS was found not to be competitive and is not considered here.
The initial guess
$\mathbf{h}_{\boldsymbol{\theta}}^{(0)}$ under Laplace scaling is
given by~(\ref{eq:laplace_guess}), whereas $\mathbf{G}_{\boldsymbol{\theta}}^{(K)}=
\mathbf{G}_{\boldsymbol{\theta}}^{(0)}$ given in~(\ref{eq:laplace_guess_G}). 
This (fixed) matrix was also used as the scaling matrix in the approximate
Newton iterations for $K=0,1,2$ (see supplementary material, Section~\ref{sec:Details-related-to}
for more details).

\subsection{Data and results}
\begin{table}[t]
\centering{}%
\begin{tabular}{lccccccc}
\hline 
 & {\scriptsize{}Stan-Prior} &  & {\scriptsize{}Stan-Laplace} &  & {\scriptsize{}Stan-Laplace} &  & {\scriptsize{}Stan-Laplace}\tabularnewline
 &  &  & {\scriptsize{}$K=0$} &  & {\scriptsize{}$K=1$} &  & {\scriptsize{}$K=2$}\tabularnewline
\cline{2-2} \cline{4-4} \cline{6-6} \cline{8-8} 
{\scriptsize{}CPU time (s)} & {\scriptsize{}6437} &  & {\scriptsize{}910} &  & {\scriptsize{}1196} &  & {\scriptsize{}1452}\tabularnewline
{\scriptsize{}$\mu_{1:5}$ ESS (min , max)} & {\scriptsize{}(832 , 918)} &  & {\scriptsize{}(967 , 1000)} &  & {\scriptsize{}(987 , 1000)} &  & {\scriptsize{}(985 , 1000)}\tabularnewline
{\scriptsize{}$\sigma_{1:5}$ ESS (min , max)} & {\scriptsize{}(301 , 349)} &  & {\scriptsize{}(1000 , 1000)} &  & {\scriptsize{}(1000 , 1000)} &  & {\scriptsize{}(1000 , 1000)}\tabularnewline
{\scriptsize{}$\delta_{1:5}$ ESS (min , max)} & {\scriptsize{}(357 , 501)} &  & {\scriptsize{}(980 , 1000)} &  & {\scriptsize{}(986 , 1000)} &  & {\scriptsize{}(975 , 1000)}\tabularnewline
{\scriptsize{}$h_{i,j}$ ESS (min , max)} & {\scriptsize{}(972 , 1000)} &  & {\scriptsize{}(984 , 1000)} &  & {\scriptsize{}(1000 , 1000)} &  & {\scriptsize{}(1000 , 1000)}\tabularnewline
{\scriptsize{}$\nu$ ESS} & {\scriptsize{}562} &  & {\scriptsize{}1000} &  & {\scriptsize{}1000} &  & {\scriptsize{}1000}\tabularnewline
{\scriptsize{}$x_{1:5,1}$ ESS (min , max)} & {\scriptsize{}(986 , 1000)} &  & {\scriptsize{}(1000 , 1000)} &  & {\scriptsize{}(1000 , 1000)} &  & {\scriptsize{}(1000 , 1000)}\tabularnewline
{\scriptsize{}$u_{1:5,1}$ ESS (min , max)} & {\scriptsize{}(871 , 959)} &  & {\scriptsize{}(1000 , 1000)} &  & {\scriptsize{}(1000 , 1000)} &  & {\scriptsize{}(1000 , 1000)}\tabularnewline
\hline 
\end{tabular}\caption{\label{tab:Effective-sample-sizes-Wishart}Effective sample sizes
and CPU times for the inverse Wishart model~(\ref{eq:wishart_mod_1}-\ref{eq:wishart_mod_3}).
The parameters are grouped, and the reported ESS figures are (min,
max) across each group. All of the results are averages across 8 independent
replica of each experiment. Here, $u_{s,1}$ is the first element
in $\mathbf{u}_{s}$. Under Prior transport map, $u_{1:5,1}$ is identical
to $\eta_{1:5,1}$ in~(\ref{eq:wishart_mod_3}).}
\end{table}

\begin{figure}[t]
\begin{centering}
\includegraphics[scale=0.5]{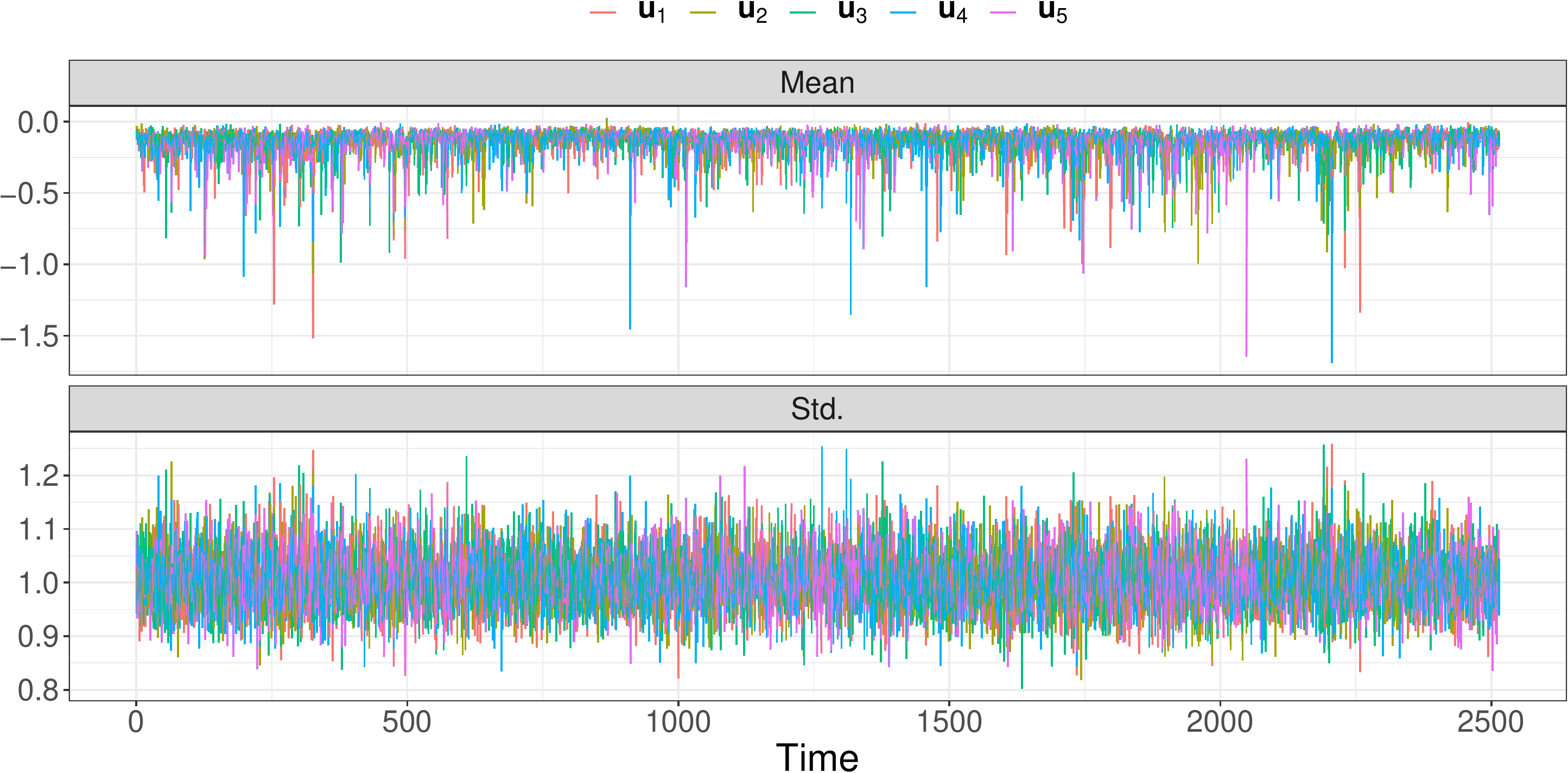}
\caption{\label{fig:Posterior-mean-and-wishart}
Posterior mean and standard
deviation of $\mathbf{u}_{s},\;s=1,\dots,5$, for the inverse Wishart
model~(\ref{eq:wishart_mod_1}-\ref{eq:wishart_mod_3}) under Laplace
transport map with $K=0$. The results are for a single representative
simulation replica with 1000 sampling iterations.}
\par\end{centering}
\end{figure}
The data set of $D=2,514$ observations of daily realized covariance
matrices of $r=5$ stocks (American Express, Citigroup, General Electric,
Home Depot, and IBM) spanning Jan. 1st, 2000 to Dec. 31, 2009 is described
in detail in \citet{Golosnoy2012}. The same model and data set was
considered in \citet{grothe2017}, where Gibbs sampling procedures
were considered. From \citet{grothe2017}, it is seen that even with
close to iid sampling from $p(\mathbf{x}_{1:r}|\boldsymbol{\theta},Y_{1:D})$,
the chains for $\nu$ and $\sigma_{s},\;s=1,\dots,r$ mix rather poorly
under Gibbs sampling. 

The ESSes for the parameters and the first elements of $\mathbf{x}_{s}$
and $\mathbf{u}_{s}$, and CPU times for Stan-Prior and Stan-Laplace
are given in Table~\ref{tab:Effective-sample-sizes-Wishart}. Corresponding
posterior means and standard deviations for Stan-Laplace ($K=0$)
are given in Table~\ref{tab:Posterior-mean-and-wishart} in the supplementary
material and these are very much in line with \citet[Table 5]{grothe2017}.

From Table~\ref{tab:Effective-sample-sizes-Wishart} it is seen that
the proposed methodology Stan-Laplace outperforms the benchmark Stan-Prior,
both in terms CPU time (the modified target is highly non-Gaussian
and thus requires many integration steps) and ESS. Indeed, Stan-Laplace
with $K=0$ is at least a full order of magnitude faster in terms
of ESS per CPU time than Stan-Prior for the ``difficult'' parameters
$\nu$ and $\sigma_{s},\;s=1,\dots,r$. The added per evaluation computational
cost of the more accurate Laplace approximations ($K=1$ and $K=2$)
is not worthwhile, and this again corroborates the finds above that
only crude location- and scale information with respect to $p(\mathbf{x}_{1:r}|\boldsymbol{\theta},Y_{1:D})$
is needed. Figure~\ref{fig:Posterior-mean-and-wishart} depicts the
posterior mean and (marginal) standard deviation of each $\mathbf{u}_{s}$,
for Stan-Laplace with $K=0$. It is seen that the posterior standard
deviations are close to 1, which one would expect in the case of close
to perfect decoupling, i.e. is indicative that any funnel effects have
been removed. The posterior means, on the other hand, are somewhat off
0, which is related both to the usage of the initial guess~(\ref{eq:laplace_guess})
and the fact that~(\ref{eq:wishart-marg-target}) is non-Gaussian
and thus cannot be exactly decoupled using a Gaussian importance density.
Figures~\ref{fig:Posterior-mean-and-wishart-2},\ref{fig:Posterior-mean-and-wishart-10}
in the supplementary material shows corresponding plots for $K=2$
and $K=10$, and it is seen that the posterior means of $\mathbf{u}_{s}$
are closer to zero, but some deviation still exact due to the non-Gaussian
target.

Comparing the computational performance to the Gibbs sampler in \citet{grothe2017},
it is seen that Stan-Laplace is also roughly an order of magnitude
faster than a Gibbs sampler. This comparison is somewhat complicated
by that \citet{grothe2017} employ parallel processing (over $s$)
when sampling the latent states $\mathbf{x}_{s}$, and that the
computations in \citet{grothe2017} are done in MATLAB, whereas Stan
is based on compiled C++ code. In this consideration, also the fact
that a model with 20 parameters and 12,570 latent variables can be
fitted using a few minutes of CPU time and minimal coding efforts
in Stan must be weighed against the typically time consuming and error-prone
development efforts to develop Gibbs samplers tailored for any given
model.

\section{Discussion \label{sec:Discussion}}

The paper proposes and evaluates importance sampler-based transport
map HMC for Bayesian hierarchical models. The methodology relies on
using off-the-shelf importance sampling strategies for high-dimensional
latent variables to construct a modified target distribution that
is easily sampled using (fixed metric) HMC. Indeed, as illustrated, the proposed
methodology can lead to large speedups relative to relevant benchmarks
for models with high-dimensional latent variables, while still being
easily implemented using e.g. Stan.

Two strategies for selecting the involved importance samplers were
considered in order to assess the optimal accuracy versus computational
cost-tradeoff. The main insight in this regard is that only rather
crude importance densities/transport maps (e.g. Laplace or DRHMC-type)
 are required when these are applied in
the present framework. This observation is very much to the contrary
to the importance sampling literature at large, where typically very
accurate importance densities are required to produce reliable approximations
to marginal likelihood functions when integrating over high-dimensional
latent variables. 

The proposed methodology, with Laplace transport maps and few or no Newton iterations
lead to similar transport maps as those used in DRHMC in the cases where DRHMC is applicable.
Thus the Laplace transport map approach may, in a rather broad sense, be seen as a generalization of DRHMC
to models with nonlinear structures where DRHMC is not applicable.

Finally, there is scope for future research in developing software that can encompass
a large class of models, and which implements the proposed methodology in a user-friendly
manner. In particular, such software should include a sparse Cholesky algorithm for more
general sparsity structures so that Laplace-based transport maps for
e.g. multivariate latent state dynamic models and
spatial models can be considered.

%\bibliographystyle{chicago}
%\bibliography{HMC_BibTex}
\putbib
\end{bibunit}
\appendix
\begin{bibunit}
\newpage{}
\noindent \begin{center}
{\LARGE{}Supplementary Material for ``Importance Sampling-based Transport
map Hamiltonian Monte Carlo for Bayesian Hierarchical Models''}{\LARGE\par}
\par\end{center}

\setcounter{page}{1}Equation numbers $<$~\ref{eq:Hamiltonian} refer
to the equations in the main text.

\section{The \citet{Lindsten2016}-integrator\label{sec:The--integrator}}

The the pseduo-marginal HMC (PM-HMC) algorithm of \citet{Lindsten2016} can be viewed as a standard HMC algorithm for simulating
the random vector $\mathbf{q}=(\boldsymbol{\theta}',\mathbf{u}')'$
from the modified target densities~(\ref{eq:t-target}) or~(\ref{eq:t-target-2}).
Proceeding with representation~(\ref{eq:t-target-2}), the Hamiltonian
is taken to be 
\begin{equation}
H(\boldsymbol{\theta},\mathbf{u},\mathbf{p}_{\boldsymbol{\theta}},\mathbf{p}_{\mathbf{u}})=-\log\omega_{\mathbf{\boldsymbol{\theta}}}(\mathbf{u})-\log p(\mathbf{\boldsymbol{\theta}})+\frac{1}{2}\mathbf{u}'\mathbf{u}+\frac{1}{2}\mathbf{p}_{\boldsymbol{\theta}}'\mathbf{M}_{\boldsymbol{\theta}}^{-1}\mathbf{p}_{\boldsymbol{\theta}}+\frac{1}{2}\mathbf{p}_{\mathbf{u}}'\mathbf{p}_{\mathbf{u}},\label{eq:Hamiltonian}
\end{equation}
where $\mathbf{p}_{\boldsymbol{\theta}}\in\mathbb{R}^{d}$ and $\mathbf{p}_{\mathbf{u}}\in\mathbb{R}^{D}$
are the artificial momentum variables specific to $\boldsymbol{\theta}$
and $\mathbf{u}$, respectively. Note that for this form of the extended
Hamiltonian the mass matrix ($\mathbf{M}$) of the compound vector
$(\boldsymbol{\theta}',\mathbf{u}')'$ is selected to be block diagonal,
where the mass matrix specific to $\boldsymbol{\theta}$ is denoted
by $\mathbf{M}_{\boldsymbol{\theta}}\in\mathbb{R}^{d\times d}$, while
the mass for $\mathbf{u}$ is set equal to the identity in order to
match the a-priori precision matrix of $\mathbf{u}$. Straight forward
modifications of~(\ref{eq:Hamiltonian}) and the proceeding theory
applies if representation~(\ref{eq:t-target}) is computationally
more convenient.

Applying Hamilton's equations~(\ref{eq:Ham_eqns}) to the extended Hamiltonian~(\ref{eq:Hamiltonian}),
for $\mathbf{q}=(\boldsymbol{\theta}',\mathbf{u}')'$ and $\mathbf{p}=(\mathbf{p}_{\boldsymbol{\theta}}',\mathbf{p}_{\mathbf{u}}')'$,
we get the following equations of motion
\begin{equation}
\frac{d}{dt}\left(\begin{array}{c}
\boldsymbol{\theta}\\
\mathbf{p_{\boldsymbol{\theta}}}\\
\mathbf{u}\\
\mathbf{p_{u}}
\end{array}\right)=\left(\begin{array}{c}
\mathbf{M}_{\boldsymbol{\theta}}^{-1}\mathbf{p}_{\boldsymbol{\theta}}\\
\nabla_{\boldsymbol{\theta}}\log p(\boldsymbol{\theta})+\nabla_{\boldsymbol{\theta}}\log\omega_{\mathbf{\boldsymbol{\theta}}}(\mathbf{u})\\
\mathbf{p}_{\mathbf{u}}\\
-\boldsymbol{u}+\nabla_{\boldsymbol{u}}\log\omega_{\mathbf{\boldsymbol{\theta}}}(\mathbf{u})
\end{array}\right).\label{eq:Ham_eqns_augmented}
\end{equation}
Equation~(\ref{eq:Ham_eqns_augmented}) shows that the Hamiltonian
transition dynamics of $(\boldsymbol{\theta},\mathbf{p}_{\boldsymbol{\theta}})$
and $(\mathbf{u},\mathbf{p}_{\mathbf{u}})$ are linked together via
their joint dependence on the importance weight $\omega_{\mathbf{\boldsymbol{\theta}}}(\mathbf{u})$.
However, this link vanishes as the MC variance of the MC estimator
$\text{{Var}}_{\mathbf{u}}[\omega_{\mathbf{\boldsymbol{\theta}}}(\mathbf{u})]$
tend to zero. In fact, an `exact' MC estimate with zero MC variance
implies that $\nabla_{\mathbf{u}}\log\omega_{\mathbf{\boldsymbol{\theta}}}(\mathbf{u})=\mathbf{0}_{D}$,
in which case the transition dynamics of $(\boldsymbol{\theta},\mathbf{p}_{\boldsymbol{\theta}})$
would be completely decoupled from that of $(\mathbf{u},\mathbf{p}_{\mathbf{u}})$
and would be (marginally) the dynamics of the `ideal' HMC algorithm
for $p(\boldsymbol{\theta}|\mathbf{y})$. Moreover, the resulting
marginal $(\mathbf{u},\mathbf{p}_{\mathbf{u}})$-dynamics would reduce
to that of a harmonic oscillator with analytical solutions given by
$\mathbf{u}(t)=\cos(t)\mathbf{u}(0)+\sin(t)\mathbf{p}_{\mathbf{u}}(0)$
and $\mathbf{p}_{\mathbf{u}}(t)=\cos(t)\mathbf{p}_{\mathbf{u}}(0)-\sin(t)\mathbf{u}(0)$.

In order to approximate the Hamiltonian transition dynamics~(\ref{eq:Ham_eqns_augmented}),
\citet{Lindsten2016} develop a symplectic integrator which for exact
likelihood estimates produces exact simulations for the dynamics of
$(\mathbf{u},\mathbf{p}_{\mathbf{u}})$ and reduces for $(\boldsymbol{\theta},\mathbf{p}_{\boldsymbol{\theta}})$
to the conventional leapfrog integrator. They derive this integrator
for the special case where the mass matrix $\mathbf{M}_{\boldsymbol{\theta}}$,
in~(\ref{eq:Hamiltonian}) and~(\ref{eq:Ham_eqns_augmented}) is restricted
to be the identity. For the more general case with an unrestricted
$\mathbf{M}_{\boldsymbol{\theta}}$ this integrator for approximately
advancing the dynamics from time $t=0$ to time $t=\varepsilon$ is given by
\begin{align}
\boldsymbol{\theta}(\varepsilon/2) & =\boldsymbol{\theta}(0)+(\varepsilon/2)\mathbf{M}_{\boldsymbol{\theta}}^{-1}\mathbf{p}_{\boldsymbol{\theta}}(0),\label{eq:int_1}\\
\mathbf{u}(\varepsilon/2) & =\cos(\varepsilon/2)\mathbf{u}(0)+\sin(\varepsilon/2)\mathbf{p}_{\mathbf{u}}(0),\\
\mathbf{p}_{\mathbf{u}}^{*} & =\cos(\varepsilon/2)\mathbf{p}_{\mathbf{u}}(0)-\sin(\varepsilon/2)\mathbf{u}(0),\\
\mathbf{p}_{\mathbf{u}}^{**} & =\mathbf{p}_{\mathbf{u}}^{*}+\varepsilon\,\nabla_{\mathbf{u}}\big\{\log\omega_{\boldsymbol{\theta}(\varepsilon/2)}(\mathbf{u}(\varepsilon/2))\big\},\label{eq:P-U}\\
\mathbf{p}_{\boldsymbol{\theta}}(\varepsilon) & =\mathbf{p}_{\theta}(0)+\varepsilon\,\nabla_{\boldsymbol{\theta}}\big\{\log p\big[\boldsymbol{\theta}(\varepsilon/2)\big]+\log\omega_{\boldsymbol{\theta}(\varepsilon/2)}(\mathbf{u}(\varepsilon/2))\big\},\label{eq:P-theta}\\
\boldsymbol{\theta}(\varepsilon) & =\boldsymbol{\theta}(\varepsilon/2)+(\varepsilon/2)\mathbf{M}_{\boldsymbol{\theta}}^{-1}\mathbf{p}_{\boldsymbol{\theta}}(\varepsilon),\\
\mathbf{u}(\varepsilon) & =\cos(\varepsilon/2)\mathbf{u}(\varepsilon/2)+\sin(\varepsilon/2)\mathbf{p}_{\mathbf{u}}^{**},\\
\mathbf{p}_{\mathbf{u}}(\varepsilon) & =\cos(\varepsilon/2)\mathbf{p}_{\mathbf{u}}^{**}-\sin(\varepsilon/2)\mathbf{u}(\varepsilon/2).\label{eq:int_last}
\end{align}

\section{The EIS principle\label{sec:The-EIS-principle}}

In order to minimize the variance of IS estimates for the likelihood
$p(\mathbf{y}|\boldsymbol{\theta})=\int p(\mathbf{y}|\mathbf{x},\boldsymbol{\theta})p(\mathbf{x}|\boldsymbol{\theta})d\mathbf{x}$
of non-Gaussian and/or nonlinear latent variable models, EIS aims
at sequentially constructing an IS density which approximates, as
closely as possible, the (infeasible) optimal IS density $m^{*}(\mathbf{x}|\boldsymbol{\theta})\propto p(\mathbf{y}|\mathbf{x},\boldsymbol{\theta})p(\mathbf{x}|\boldsymbol{\theta})$,
which would reduce the variance of likelihood estimates to zero.

With reference to the likelihood it is assumed that the conditional
data density $p(\mathbf{y}|\mathbf{x},\boldsymbol{\theta})$ and the
prior for the latent variables $p(\mathbf{x}|\boldsymbol{\theta})$
under the latent variable model can be factorized as functions in
$\mathbf{x}=(x_{1},\ldots,x_{D})$ into 
\begin{align}
p(\mathbf{y}|\mathbf{x},\boldsymbol{\theta})=\prod_{t=1}^{D}g_{t}(x_{t},\boldsymbol{\delta}),\qquad p(\mathbf{x}|\boldsymbol{\theta})=\prod_{t=1}^{D}f_{t}(x_{t}|\mathbf{x}_{(t-1)},\boldsymbol{\delta}),\label{eq:factorized_target}
\end{align}
where $\mathbf{x}_{(t)}=(x_{1},\ldots,x_{t})$ with $\mathbf{x}_{(D)}=\mathbf{x}$
and $\boldsymbol{\delta}=(\boldsymbol{\theta},\mathbf{y})$. Such
factorizations can be found for a broad class of models, including
dynamic non-Gaussian/nonlinear state-space models for time series,
non-Gaussian/nonlinear models with a latent correlation structure
for cross-sectional data as well as static hierarchical models without
latent correlation for which $f_{t}(x_{t}|\mathbf{x}_{(t-1)},\boldsymbol{\delta})=f_{t}(x_{t},\boldsymbol{\delta})$.
E.g., variants of EIS for univariate
and multivariate linear Gaussian states subject to nonlinear measurements
are given in \citet{Liesenfeld2003,Liesenfeld2006} and for more general
nonlinear models in \citet{Kleppe201473,MOURA2014494}. EIS implementations
with more flexible IS densities such as mixture of normal distributions
are found in \citet{KLEPPE2014449}, \citet{Scharth2016133}, \citet{grothe2017},
and \citet{LiesenfeldRichard2010} use truncated normal distributions.
Applications of EIS to models with non-Markovian latent variables
for spatial data are provided in \citet{doi:10.1108/S0731-905320160000037009,doi:10.1002/jae.2534}.
In our applications we consider univariate time series models, which
is why we use $t$ to index the elements in $\mathbf{x}$ and restrict
$x_{t}$ in~(\ref{eq:factorized_target}) to be one-dimensional.

EIS-MC estimation of likelihood functions $p(\mathbf{y}|\boldsymbol{\theta})$
associated with~(\ref{eq:factorized_target}) is based upon an IS
density $m$ for $\mathbf{x}$ which is decomposed conformably with
the factorization in~(\ref{eq:factorized_target}) into 
\begin{align}
m(\mathbf{x}|\mathbf{a})=\prod_{t=1}^{D}m_{t}(x_{t}|\mathbf{x}_{(t-1)},\mathbf{a}_{t}),\label{eq:EIS_density_global}
\end{align}
with conditional densities $m_{t}$ such that 
\begin{align}
m_{t}(x_{t}|\mathbf{x}_{(t-1)},\mathbf{a}_{t})=\frac{k_{t}(\mathbf{x}_{(t)},\mathbf{a}_{t})}{\chi_{t}(\mathbf{x}_{(t-1)},\mathbf{a}_{t})},\quad\chi_{t}(\mathbf{x}_{(t-1)},\mathbf{a}_{t})=\int k_{t}(\mathbf{x}_{(t)},\mathbf{a}_{t})dx_{t},\label{eq:EIS_density_local}
\end{align}
where ${\cal K}=\{k_{t}(\cdot,\mathbf{a}_{t}),\mathbf{a}_{t}\in{\cal A}_{t}\}$
is a preselected parametric class of density kernels indexed by auxiliary
parameters $\mathbf{a}_{t}$ and with a point-wise computable integrating
factor $\chi_{t}$. As required for the proposed methodology, it is
assumed that the IS density~(\ref{eq:EIS_density_global}) can be
simulated by sequentially generating draws from the conditional densities~
(\ref{eq:EIS_density_local}) using smooth deterministic functions
$\gamma_{t}$ such that $x_{t}=\gamma_{t}(\mathbf{a}_{t},v_{t})$
for $t=1,\ldots,D$, where $v_{t}\sim N(0,1)$.

From~(\ref{eq:factorized_target})-(\ref{eq:EIS_density_local}) results
the following factorized IS representation of the likelihood: 
\begin{align}
p(\mathbf{y}|\boldsymbol{\theta})=\int\left[\chi_{1}(\mathbf{a}_{1},\boldsymbol{\delta})\prod_{t=1}^{D}\omega_{t}(\mathbf{x}_{(t)},\mathbf{a}_{(t+1)},\boldsymbol{\delta})\right]m(\mathbf{x}|\mathbf{a})d\mathbf{x},
\end{align}
where the period-$t$ IS weight is given by 
\begin{align}
\omega_{t}(\mathbf{x}_{(t)},\mathbf{a}_{(t+1)},\boldsymbol{\delta})=\frac{g_{t}(x_{t},\boldsymbol{\delta})f_{t}(x_{t}|\mathbf{x}_{(t-1)},\boldsymbol{\delta})\chi_{t+1}(\mathbf{x}_{(t)},\mathbf{a}_{t+1},\boldsymbol{\delta})}{k_{t}(\mathbf{x}_{(t)},\mathbf{a}_{t})},\label{eq:period_t_IS_weight}
\end{align}
with $\chi_{D+1}(\cdot)\equiv1$. For any given $\mathbf{a}=(\mathbf{a}_{1},\ldots,\mathbf{a}_{D})\in{\cal A}=\times_{t=1}^{D}{\cal A}_{t}$,
the corresponding MC likelihood estimate is given by
\begin{align}
\hat{p}(\mathbf{y}|\boldsymbol{\theta},\mathbf{u})=\omega(\mathbf{x},\mathbf{a}),\qquad\omega(\mathbf{x},\mathbf{a})=\prod_{t=1}^{D}\omega_{t}(\mathbf{x}_{(t)},\mathbf{a}_{(t+1)}),\label{eq:EIS_estimate}
\end{align}
where $\mathbf{x}$ is a draw simulated from the sequential IS density
$m(\mathbf{x}|\mathbf{a})$ in~(\ref{eq:EIS_density_global}) (which
is obtained by transforming $\mathbf{u}$ using the sequence of smooth
deterministic functions $\gamma_{t}$).

In order to minimize the MC variance of the likelihood estimate~(\ref{eq:EIS_estimate}),
EIS aims at selecting values for the auxiliary parameters $\mathbf{a}$
that minimize period-by-period the MC variance of the IS weights $\omega_{t}$
in~(\ref{eq:period_t_IS_weight}) with respect to $m(\mathbf{x}|\mathbf{a})$.
This requires that the kernels $k_{t}(\mathbf{x}_{(t)},\mathbf{a}_{t})$
as functions in $\mathbf{x}_{(t)}$ provide the best possible fit
to the products $g_{t}(x_{t},\boldsymbol{\delta})f_{t}(x_{t}|\mathbf{x}_{(t-1)},\boldsymbol{\delta})\chi_{t+1}(\mathbf{x}_{(t)},\mathbf{a}_{t+1})$.
For an approximate solution to this minimization problem under the
preselected class of kernels ${\cal K}$, EIS solves the following
back-recursive sequence of least squares (LS) approximation problems:
\begin{equation}
\begin{aligned}(\hat{c}_{t},\hat{\mathbf{a}}_{t})=\arg\min_{c_{t}\in\mathbb{R},\mathbf{a}_{t}\in{\cal A}_{t}}\sum_{i=1}^{r}\Bigg\{ & \log\left[g_{t}\big(x_{t}^{(i)},\boldsymbol{\delta}\big)\;f_{t}\big(x_{t}^{(i)}|\mathbf{x}_{(t-1)}^{(i)},\boldsymbol{\delta}\big)\;\chi_{t+1}\big(\mathbf{x}_{(t)}^{(i)},\hat{\mathbf{a}}_{t+1}\big)\right]\\
 & -c_{t}-\log k_{t}\big(\mathbf{x}_{(t)}^{(i)},\mathbf{a}_{t}\big)\Bigg\}^{2},\qquad t=D,D-1,\ldots,1,
\end{aligned}
\label{eq:eis_min}
\end{equation}
where $c_{t}$ represents an intercept, and $\{{\mathbf{x}}^{(i)}\}_{i=1}^{r}$
denote $r$ iid draws simulated from $m(\mathbf{x}|\mathbf{a})$ itself.
Thus, the EIS-optimal values for the auxiliary parameters $\hat{\mathbf{a}}$
result as a fixed-point solution to the sequence $\{\hat{\mathbf{a}}^{[0]},\hat{\mathbf{a}}^{[1]},\ldots\}$
in which $\hat{\mathbf{a}}^{[j]}$ is given by~(\ref{eq:eis_min})
under draws from $m(\mathbf{x}|\hat{\mathbf{a}}^{[j-1]})$. In order
to ensure convergence to a fixed-point solution it is critical that
all the $\mathbf{x}$ draws simulated for the sequence $\{\hat{\mathbf{a}}^{[j]}\}$
be generated by using the smooth deterministic functions $\gamma_{t}$
to transform a \textsl{single set} of $rD$ Common Random Numbers
(CRNs), say $\mathbf{z}\sim N(\mathbf{0}_{rD},\mathbf{I}_{rD})$.
To initialize the fixed-point iterations $j=0,\ldots,J$, the starting
value $\hat{\mathbf{a}}^{[0]}$ can be found, e.g., from an analytical
local approximation (such as Laplace) of the EIS targets $\ln(g_{t}f_{t}\chi_{t+1})$
in~(\ref{eq:eis_min}). Convergence of the iterations to a fixed-point
solution is typically fast to the effect that a value for the number
of iterations $J$ between 2 and 4 often suffices to produce a (close
to) optimal solution \citep{Richard2007}. The MC-EIS likelihood estimate,
for a given $\boldsymbol{\theta}$, is then calculated by substituting in~
(\ref{eq:EIS_estimate}) the EIS-optimal value $\hat{\mathbf{a}}$
for $\mathbf{a}$. In order to highlight its dependence on $\boldsymbol{\theta}$
and $\mathbf{z}$ we shall use $\hat{\mathbf{a}}=\mathbf{a}(\boldsymbol{\theta},\mathbf{z})$
to denote the EIS-optimal value.

The selection of the parametric class ${\cal K}$ of EIS density kernels
$k_{t}$ is inherently specific to the latent variable model under
consideration as those kernels are meant to provide a functional approximation
in $\mathbf{x}_{(t)}$ to the product $g_{t}f_{t}\chi_{t+1}$. In
the applications below, we consider models with data densities $g_{t}$
which are log-concave in $x_{t}$ and Gaussian conditional densities
for $x_{t}$ with a Markovian structure so that $f_{t}(x_{t}|\mathbf{x}_{(t-1)},\boldsymbol{\delta})=f_{t}(x_{t}|x_{t-1},\boldsymbol{\delta})$.
This suggests selection of the $k_{t}$'s as Gaussian kernels and to
exploit that such kernels are closed under multiplication in order
to construct the $k_{t}$'s as the following parametric extensions
of the prior densities $f_{t}$: 
\begin{align}
k_{t}(x_{t},x_{t-1},\mathbf{a}_{t})=f_{t}(x_{t}|x_{t-1},\boldsymbol{\delta})\xi_{t}(x_{t},\mathbf{a}_{t}),
\end{align}
where $\xi_{t}$ is a Gaussian kernel in $x_{t}$ of the form $\xi_{t}(x_{t},\mathbf{a}_{t})=\exp\{a_{1t}x_{t}+a_{2t}x_{t}^{2}\}$
with $\mathbf{a}_{t}=(a_{1t},a_{2t})$. In this case the EIS approximation
problems~(\ref{eq:eis_min}) take the form of simple \textsl{linear}
LS-problems where $\log[g_{t}(x_{t}^{(i)},\boldsymbol{\delta})$ $\chi_{t+1}(x_{t}^{(i)},\hat{\mathbf{a}}_{t+1})]$
are regressed on a constant, $x_{t}^{(i)}$ and $[x_{t}^{(i)}]^{2}$.
In fact,~(\ref{eq:eis_min}) reduces to linear LS regressions for all kernels $k_{t}$ chosen within the exponential family \citep{Richard2007}, which simplifies implementation. However, it is important
to note that EIS is by no means restricted to the use of IS densities
from the exponential family nor to models with low-order Markovian
specifications for the latent variables. 

The EIS approach as outlined above differs from standard IS in that
it uses IS densities whose parameters $\hat{\mathbf{a}}=\mathbf{a}(\boldsymbol{\theta},\mathbf{z})$
are (conditional on $\boldsymbol{\theta}$) random variables as they
depend via the EIS fixed-point repressions~(\ref{eq:eis_min}) on
the CRNs $\mathbf{z}$. This calls for specific rules for implementing
EIS which ensure that the resulting MC likelihood estimates meet the
qualifications needed for their use within PM-HMC. In order to ensure
that the EIS likelihood estimate~(\ref{eq:EIS_estimate}) based on
the random numbers $\mathbf{u}$ is unbiased the latter need to be
a set of random draws different from the CRNs $\mathbf{z}$ used to
find $\hat{\mathbf{a}}$ \citep{KLEPPE2014449}. Note also that since
$\hat{\mathbf{a}}$ is an implicit function of $\boldsymbol{\theta}$,
maximal accuracy requires us to rerun the EIS fixed-point regressions
for any new value of $\boldsymbol{\theta}$. In order to ensure that
the resulting EIS likelihood estimate~(\ref{eq:EIS_estimate}) as
a function of $\hat{\mathbf{a}}$ is smooth in $\boldsymbol{\theta}$,
$\hat{\mathbf{a}}$ itself needs to be a smooth function of $\boldsymbol{\theta}$.
This can be achieved by presetting the number of fixed-point iterations
$J$ across all $\boldsymbol{\theta}$-values to a fixed number, rather
than using a stopping rule based on a relative-change threshold.

The EIS-specific tuning parameters are the number of $\mathbf{x}^{(i)}$-draws
$r$ used to run the EIS optimization process, the number of fixed-point
iterations on the EIS regressions $J$, and the number of $\mathbf{x}^{(i)}$-draws
$n$ for the likelihood estimate~(\ref{eq:EIS_estimate}). Those parameters
should be selected to balance the trade-off between EIS computing
time and the quality of the resulting EIS density with respect to
the MC accuracy. In particular, for $r$ it is recommended to select
it as small as possible while retaining the EIS fixed-point regressions
numerically stable and the parameter $J$ should be set such that
it is guaranteed that the fixed-point sequence $\{\mathbf{a}^{[j]}\}_{j}$
approximately converge for the $\boldsymbol{\theta}$ values in the
relevant range of the parameter space. In our applications, where
the selected class of kernels ${\cal K}$ imply that the EIS regressions
are linear in the EIS parameters $\mathbf{a}_{t}$, we find that a
$J$ set equal to 1 or 2 and an $r$ about 2 times the number of parameters
in $(\mathbf{a}_{t},c_{t})$ suffice. We obtain EIS kernels $k_{t}$
providing highly accurate approximations to the targeted product $g_{t}f_{t}\chi_{t+1}$,
with an $R^{2}$ of the EIS regressions in the final iteration typically
larger than 0.95. 

\section{Details related to the example models in Section \ref{sec:Simulation-study}\label{sec:Details-related-to-sim-study}}
\subsection{SV model}
For the SV model, the standard prior assumptions for the parameters
$\boldsymbol{\theta}=(\gamma,\delta,\nu)$ are the following: for
$\gamma$ we use a flat prior, for $(\delta+1)/2$ a Beta prior ${\cal {B}(\alpha,\beta)}$
with $\alpha=20$ and $\beta=1.5$, and for $\nu^{2}$ a scaled inverted-$\chi^{2}$
prior $p_{0}s_{0}/\chi_{(p_{0})}^{2}$ with $p_{0}=10$ and $s_{0}=0.01$.
For numerical stability we use the parametrization $\boldsymbol{\theta}^{*}=(\gamma,\mbox{arctanh}\,\delta,\log\nu^{2})$
together with the priors for $\boldsymbol{\theta}^{*}$ to run the HMC algorithms, where the priors are derived from
those on $\boldsymbol{\theta}$.

For the Laplace transport map, $\mathbf G_{\boldsymbol \theta}^{(0)}$ and $\mathbf h_{\boldsymbol \theta}^{(0)}$
are taken to be identical to~(\ref{eq:laplace_guess_G},\ref{eq:laplace_guess}). More refined solutions are 
found using Newton iterations;
\begin{align*}
\mathbf{h}_{\boldsymbol{\theta}}^{(k)} & =\mathbf{h}_{\boldsymbol{\theta}}^{(k-1)}+
\left[\nabla_{\mathbf{x}}^{2}\log\left[p(\mathbf{x}|\boldsymbol{\theta})p(\mathbf{y}|\mathbf{x},\boldsymbol{\theta})\right]_{\mathbf{x}=\mathbf{h}_{\boldsymbol{\theta}}^{(k-1)}}\right]^{-1}\left\{ \nabla_{\mathbf{x}}\log\left[p(\mathbf{x}|\boldsymbol{\theta})p(\mathbf{y}|\mathbf{x},\boldsymbol{\theta})\right]
_{\mathbf{x}=\mathbf{h}_{\boldsymbol{\theta}}^{(k-1)}}\right\} ,\\
\mathbf{G}_{\boldsymbol{\theta}}^{(k)} & =
\nabla_{\mathbf{x}}^{2}\log\left[p(\mathbf{x}|\boldsymbol{\theta})p(\mathbf{y}|\mathbf{x},\boldsymbol{\theta})\right]_{\mathbf{x}=\mathbf{h}_{\boldsymbol{\theta}}^{(k-1)}}.
\end{align*}
for $k=1,2,\dots,K$.
Further modifications, including changing to $\mathbf{G}_{\boldsymbol{\theta}}^{(k)}  =
\nabla_{\mathbf{x}}^{2}\log\left[p(\mathbf{x}|\boldsymbol{\theta})p(\mathbf{y}|\mathbf{x},\boldsymbol{\theta})\right]_{\mathbf{x}=\mathbf{h}_{\boldsymbol{\theta}}^{(k)}}$ (at the
cost of one additional Cholesky factorization), or keeping 
$\mathbf{G}_{\boldsymbol{\theta}}^{(k)} = \mathbf{G}_{\boldsymbol{\theta}}^{(0)}$ 
(costs only a single Cholesky factorization)
both in the transport map and as the scaling matrix in the 
Newton iterations was tried, but did not produce better results.

It is straight forward to show that $\mathbf G_{\boldsymbol \theta, \mathbf y | \mathbf x}
= 0.5 \mathbf I_D$ is also the Fisher information of $p(\mathbf y | \mathbf x)$ with respect to $\mathbf x$ (i.e.
$p(\mathbf y | \mathbf x)$ is a constant information parameterization). Hence
also Stan-Laplace $K=0$ may be interpreted as a special case of DRHMC \citep{1806.02068}. 

\subsection{Gamma model}
For the Gamma model, the priors on the parameters $\boldsymbol{\theta}=(\tau,\beta,\delta,\nu)$
are as follows; we use flat priors for $\log\tau$ as well as $\log\beta$,
a Beta ${\cal {B}(\alpha,\beta)}$ with $\alpha=20$ and $\beta=1.5$
for $(\delta+1)/2$, and a scaled inverted-$\chi^{2}$ for $\nu^{2}$
with $p_{0}s_{0}/\chi_{(p_{0})}^{2}$ and $p_{0}=10$, $s_{0}=0.01$.
For the LD computations we use the parameterization 
$\boldsymbol{\theta}^{*}=(\log\tau,\log\beta,\mbox{arctanh}\,\delta,\log\nu^{2})$.

For this model, the same strategy for calculating the Laplace transport map as for the SV
model was used. Notice that here $\mathbf G_{\boldsymbol \theta, \mathbf y | \mathbf x} = 
\tau^{-1}\mathbf I_D$ is also the Fisher information of $p(\mathbf y | \mathbf x)$ with respect to $\mathbf x$.
Hence, Stan-Laplace, $K=0$ may be interpreted as a DRHMC method.

\subsection{CEV model}
For the CEV model, for $\alpha$ and $\beta$ we assume Gaussian priors
both with $N(0,1000)$, for $\gamma$ a uniform prior on the interval
$[0,4]$, and for $\sigma_{x}^{2}$ and $\sigma_{y}^{2}$ uninformative
inverted-$\chi^{2}$ priors with $p(\sigma_{x}^{2})\propto1/\sigma_{x}^{2}$
and $p(\sigma_{y}^{2})\propto1/\sigma_{y}^{2}$. The LD computations
are conducted on the following transformed parameters: $\boldsymbol{\theta}^{*}=(\alpha,\beta,\gamma,\log{\sigma_{x}^{2}},\log{\sigma_{y}^{2}})$.

For the CEV model, the precision of the latent state prior is does not have closed-form, which
precludes the application of~(\ref{eq:laplace_guess_G},\ref{eq:laplace_guess}). However,
it is known that the measurement densities has a very small variance, hence 
$\mathbf h_{\boldsymbol \theta}^{(0)} = \mathbf y$ seems sensible. Subsequently, a full Newton
iteration is performed:
\begin{align*}
\mathbf{h}_{\boldsymbol{\theta}}^{(k)} & =\mathbf{h}_{\boldsymbol{\theta}}^{(k-1)}+
\left[\nabla_{\mathbf{x}}^{2}\log\left[p(\mathbf{x}|\boldsymbol{\theta})p(\mathbf{y}|\mathbf{x},\boldsymbol{\theta})\right]_{\mathbf{x}=\mathbf{h}_{\boldsymbol{\theta}}^{(k-1)}}\right]^{-1}\left\{ \nabla_{\mathbf{x}}\log\left[p(\mathbf{x}|\boldsymbol{\theta})p(\mathbf{y}|\mathbf{x},\boldsymbol{\theta})\right]
_{\mathbf{x}=\mathbf{h}_{\boldsymbol{\theta}}^{(k-1)}}\right\} ,\\
\mathbf{G}_{\boldsymbol{\theta}}^{(k)} & =
\nabla_{\mathbf{x}}^{2}\log\left[p(\mathbf{x}|\boldsymbol{\theta})p(\mathbf{y}|\mathbf{x},\boldsymbol{\theta})\right]_{\mathbf{x}=\mathbf{h}_{\boldsymbol{\theta}}^{(k-1)}}.
\end{align*}
for $k=1,2,\dots,K$. Further modifications, including changing to $\mathbf{G}_{\boldsymbol{\theta}}^{(k)}  =
\nabla_{\mathbf{x}}^{2}\log\left[p(\mathbf{x}|\boldsymbol{\theta})p(\mathbf{y}|\mathbf{x},\boldsymbol{\theta})\right]_{\mathbf{x}=\mathbf{h}_{\boldsymbol{\theta}}^{(k)}}$ (at the
cost of one additional Cholesky factorization) did not improve the fit sufficiently to warrant the 
additional computation.

\section{Details related to the realized volatility model in Section \ref{sec:Realistic-application}\label{sec:Details-related-to}}

\begin{table}
\centering{}{\scriptsize{}}%
\begin{tabular}{lcccccccccc}
\hline 
 & {\scriptsize{}$\mu_{1}$} & {\scriptsize{}$\mu_{2}$} & {\scriptsize{}$\mu_{3}$} & {\scriptsize{}$\mu_{4}$} & {\scriptsize{}$\mu_{5}$} & {\scriptsize{}$\delta_{1}$} & {\scriptsize{}$\delta_{2}$} & {\scriptsize{}$\delta_{3}$} & {\scriptsize{}$\delta_{4}$} & {\scriptsize{}$\delta_{5}$}\tabularnewline
\hline 
{\scriptsize{}post. mean} & {\scriptsize{}4.16} & {\scriptsize{}4.12} & {\scriptsize{}3.72} & {\scriptsize{}4.11} & {\scriptsize{}3.53} & {\scriptsize{}0.97} & {\scriptsize{}0.98} & {\scriptsize{}0.96} & {\scriptsize{}0.94} & {\scriptsize{}0.96}\tabularnewline
{\scriptsize{}post. std.} & {\scriptsize{}0.2} & {\scriptsize{}0.25} & {\scriptsize{}0.15} & {\scriptsize{}0.1} & {\scriptsize{}0.13} & {\scriptsize{}0.005} & {\scriptsize{}0.004} & {\scriptsize{}0.006} & {\scriptsize{}0.008} & {\scriptsize{}0.006}\tabularnewline
\hline 
 & {\scriptsize{}$\sigma_{1}$} & {\scriptsize{}$\sigma_{2}$} & {\scriptsize{}$\sigma_{3}$} & {\scriptsize{}$\sigma_{4}$} & {\scriptsize{}$\sigma_{5}$} & {\scriptsize{}$\nu$} &  &  &  & \tabularnewline
\hline 
{\scriptsize{}post. mean} & {\scriptsize{}0.31} & {\scriptsize{}0.26} & {\scriptsize{}0.29} & {\scriptsize{}0.28} & {\scriptsize{}0.25} & {\scriptsize{}33.61} &  &  &  & \tabularnewline
{\scriptsize{}post. std.} & {\scriptsize{}0.009} & {\scriptsize{}0.008} & {\scriptsize{}0.009} & {\scriptsize{}0.009} & {\scriptsize{}0.009} & {\scriptsize{}0.283} &  &  &  & \tabularnewline
\hline 
 & {\scriptsize{}$h_{2,1}$} & {\scriptsize{}$h_{3,1}$} & {\scriptsize{}$h_{4,1}$} & {\scriptsize{}$h_{5,1}$} & {\scriptsize{}$h_{3,2}$} & {\scriptsize{}$h_{4,2}$} & {\scriptsize{}$h_{5,2}$} & {\scriptsize{}$h_{4,3}$} & {\scriptsize{}$h_{5,3}$} & {\scriptsize{}$h_{5,4}$}\tabularnewline
\hline 
{\scriptsize{}post. mean} & {\scriptsize{}0.39} & {\scriptsize{}0.29} & {\scriptsize{}0.29} & {\scriptsize{}0.23} & {\scriptsize{}0.20} & {\scriptsize{}0.17} & {\scriptsize{}0.12} & {\scriptsize{}0.22} & {\scriptsize{}0.18} & {\scriptsize{}0.11}\tabularnewline
{\scriptsize{}post. std.} & {\scriptsize{}0.003} & {\scriptsize{}0.003} & {\scriptsize{}0.003} & {\scriptsize{}0.002} & {\scriptsize{}0.003} & {\scriptsize{}0.003} & {\scriptsize{}0.002} & {\scriptsize{}0.004} & {\scriptsize{}0.003} & {\scriptsize{}0.002}\tabularnewline
\hline 
 & {\scriptsize{}$x_{1,1}$} & {\scriptsize{}$x_{2,1}$} & {\scriptsize{}$x_{3,1}$} & {\scriptsize{}$x_{4,1}$} & {\scriptsize{}$x_{5,1}$} & {\scriptsize{}$u_{1,1}$} & {\scriptsize{}$u_{2,1}$} & {\scriptsize{}$u_{3,1}$} & {\scriptsize{}$u_{4,1}$} & {\scriptsize{}$u_{5,1}$}\tabularnewline
\hline 
{\scriptsize{}post. mean} & {\scriptsize{}5.23} & {\scriptsize{}5.28} & {\scriptsize{}4.27} & {\scriptsize{}5.46} & {\scriptsize{}5.11} & {\scriptsize{}-0.08} & {\scriptsize{}-0.05} & {\scriptsize{}-0.07} & {\scriptsize{}-0.10} & {\scriptsize{}-0.10}\tabularnewline
{\scriptsize{}post. std.} & {\scriptsize{}0.206} & {\scriptsize{}0.195} & {\scriptsize{}0.198} & {\scriptsize{}0.205} & {\scriptsize{}0.202} & {\scriptsize{}1.022} & {\scriptsize{}0.993} & {\scriptsize{}0.99} & {\scriptsize{}1.031} & {\scriptsize{}1.049}\tabularnewline
\hline 
\end{tabular}\caption{\label{tab:Posterior-mean-and-wishart}Posterior mean and standard
deviations for the inverse Wishart model~(\ref{eq:wishart_mod_1}-\ref{eq:wishart_mod_3})
based on Stan-Laplace, $K=0$. All figures are means across 8 independent
replica. Here, $u_{s,1}$ is the first element in $\mathbf{u}_{s}$,
and should be close to standard normal when the transport map produces
a sufficient de-coupling effect.}
\end{table}
\begin{figure}
\begin{centering}
\includegraphics[scale=0.5]{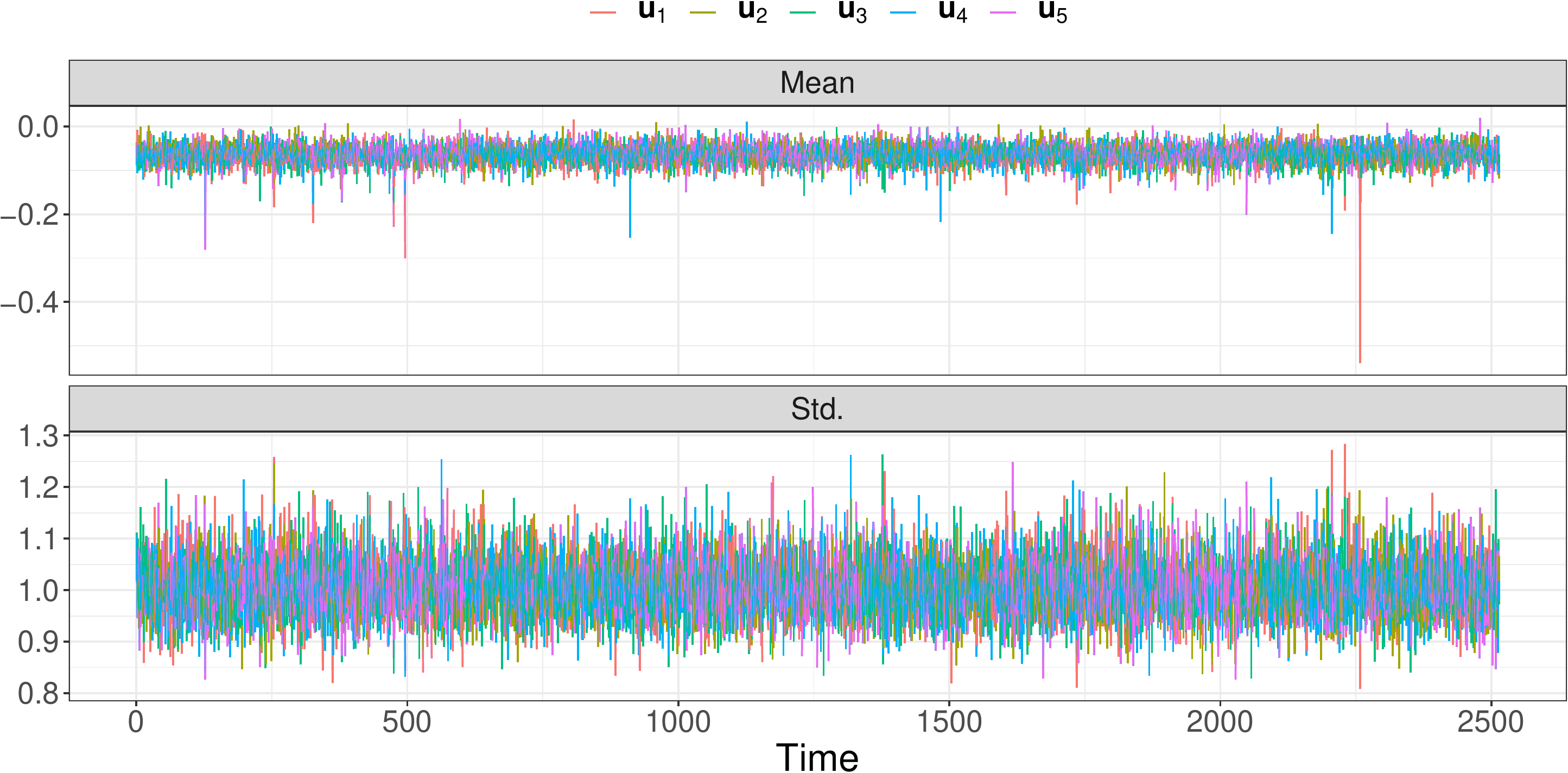}\caption{\label{fig:Posterior-mean-and-wishart-2}Posterior mean and standard
deviation of $\mathbf{u}_{s},\;s=1,\dots,5$, for the inverse Wishart
model~(\ref{eq:wishart_mod_1}-\ref{eq:wishart_mod_3}) under Laplace
transport map with $K=2$. }
\par\end{centering}
\end{figure}
\begin{figure}
\centering{}\includegraphics[scale=0.5]{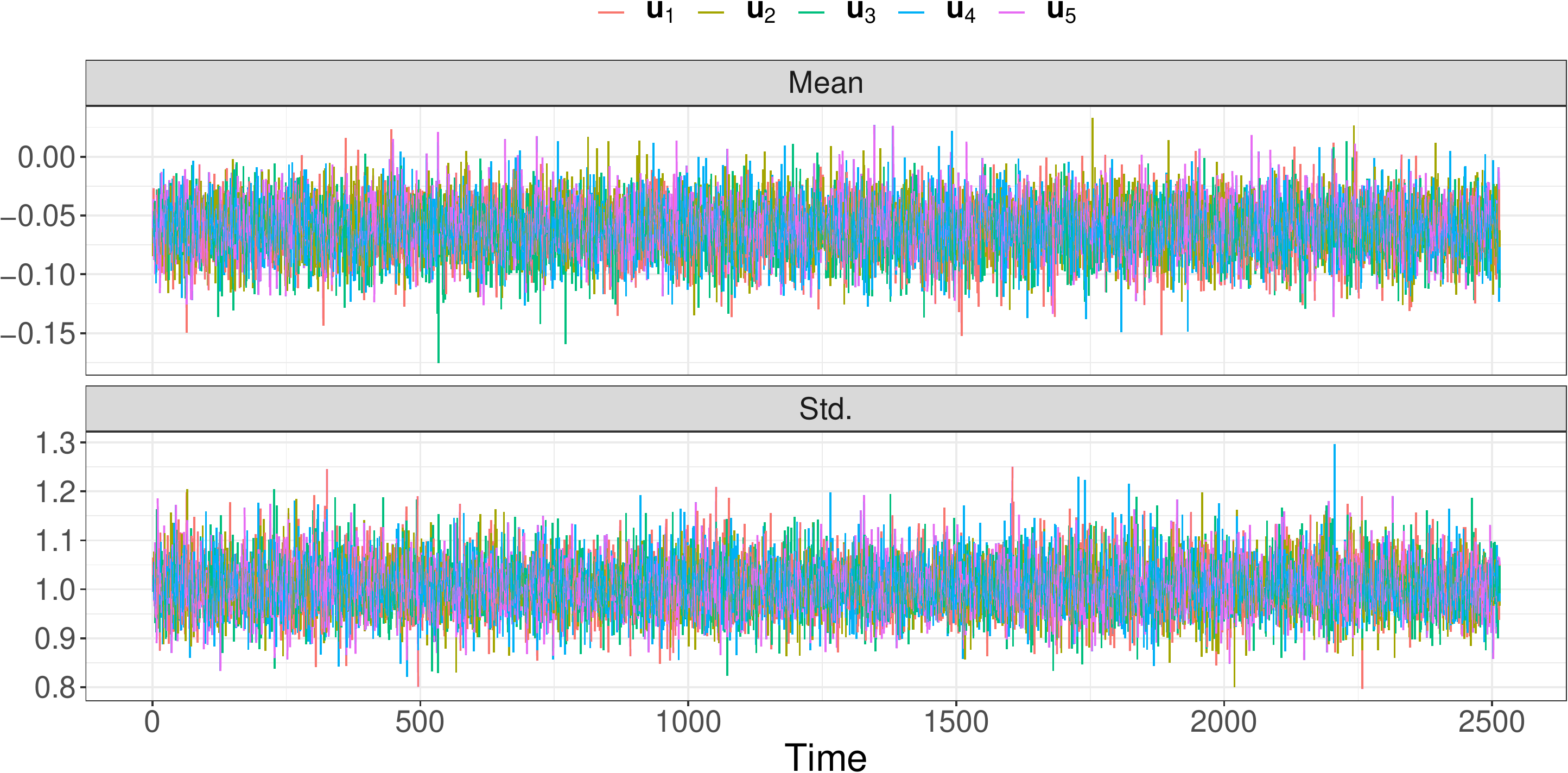}\caption{\label{fig:Posterior-mean-and-wishart-10}Posterior mean and standard
deviation of $\mathbf{u}_{s},\;s=1,\dots,5$, for the inverse Wishart
model~(\ref{eq:wishart_mod_1}-\ref{eq:wishart_mod_3}) under Laplace
transport with $K=10$.}
\end{figure}
The (normalized) observation density is given by: 
\[
p(\mathbf{Y}_{t}|\Sigma_{t},\nu)=\frac{|\boldsymbol{\Sigma}_{t}|^{\frac{\nu}{2}}}{2^{\frac{\nu r}{2}}\pi^{\frac{r(r-1)}{4}}\prod_{s=1}^{r}\Gamma\left([\nu+1-s]/2\right)}|\mathbf{Y}_{t}|^{-\frac{\nu+r+1}{2}}\exp\left(-\frac{1}{2}\text{tr}\left[\boldsymbol{\Sigma}_{t}\mathbf{Y}_{t}^{-1}\right]\right).
\]
In the Stan implementation, $\prod_{t=1}^{D}|\mathbf{Y}_{t}|$ and
$\mathbf{Y}_{t}^{-1},\;t=1,\dots,D$ where precomputed. 

The (independent) priors used to complete the model specification
in Section~\ref{subsec:Wishart-Model} are as follows: $\mu_{s}\sim N(0,25)$,
$\delta_{s}\sim\text{uniform}(-1,1)$, $\sigma_{s}^{2}\sim p_{0}s_{0}/\chi_{p_{0}}^{2}$
where $p_{0}=4$ and $s_{0}=0.25$, $h_{i,j}\sim N(0,100)$. Finally,
a flat prior on $(6.0,\infty)$ was chosen for $\nu$. 

Posterior- means and standard deviations of the parameters and the
first elements in $\mathbf{x}_{s}$ and $\mathbf{u}_{s}$ are given
in Table~\ref{tab:Posterior-mean-and-wishart}. The results are very
much in line with those of \citet{grothe2017}. 

The Laplace transport maps for each of $\mathbf x_s,\;s=1,\dots,r$ are constructed as follows;
the initial guesses for $\mathbf{h}_{\boldsymbol{\theta}}^{(0)}$ and $\mathbf{G}_{\boldsymbol{\theta}}^{(0)}$ 
are those given in~(\ref{eq:laplace_guess_G},\ref{eq:laplace_guess}), applied to~(\ref{eq:wishart-marg-target}).
The mean is further refined via the following approximate Newton iteration
\begin{equation*}
\mathbf{h}_{\boldsymbol{\theta}}^{(k)}  =\mathbf{h}_{\boldsymbol{\theta}}^{(k-1)}+
\left[\mathbf{G}_{\boldsymbol{\theta}}^{(0)}\right]^{-1}\left\{ \nabla_{\mathbf{x}}\log\left[p(\mathbf{x}|\boldsymbol{\theta})p(\mathbf{y}|\mathbf{x},\boldsymbol{\theta})\right]
_{\mathbf{x}=\mathbf{h}_{\boldsymbol{\theta}}^{(k-1)}}\right\} ,
\end{equation*}
whereas $\mathbf{G}_{\boldsymbol{\theta}}^{(k)}=\mathbf{G}_{\boldsymbol{\theta}}^{(0)}$ is kept
fixed which result in that only a single Cholesky factorization is required.
Figures~\ref{fig:Posterior-mean-and-wishart-2},\ref{fig:Posterior-mean-and-wishart-10}
show the posterior mean and standard deviations of $\mathbf{u}_{s}$
over time $t$ for Stan-Laplace, $K=2$ and $K=10$ respectively. It is seen that even with
the approximate Newton iteration, the iteration makes $\mathbf u_s$ have a mean close to
zero, where the remaining deviation from zero for $K=10$ iterations in Figure~\ref{fig:Posterior-mean-and-wishart-10}
is presumably due to the non-quadratic nature of the "measurement density" in~(\ref{eq:wishart-marg-target})
(in addition to Monte Carlo variation).

\putbib
\end{bibunit}
\end{document}